\documentclass{article}
\usepackage{arxiv}
\usepackage{hyperref}
\usepackage{graphicx}
\usepackage{bm}
\usepackage{amsmath}
\usepackage{amssymb}
\usepackage{mathtools}
\usepackage{xcolor}
\usepackage{caption}
\usepackage{subcaption}
\usepackage{tikz}
\usepackage{comment}
\usepackage{amsmath,amssymb,amsthm}
\usepackage{bm}
\usepackage{xcolor}
\usepackage[normalem]{ulem}
\newcommand{\COMMENT}[1]{\textcolor{cyan}{{[ \sc{#1} ]}}} 

\newcommand{\TODO}[1]{\textcolor{magenta}{\bf TODO {#1}}}

\newlength{\figwidth}
\newlength{\figwidthtwo}
\newlength{\figwidththree}
\newcommand{\fref}[1]{Fig.\,\ref{#1}}

\newcommand{\eref}[1]{Eq.\,(\ref{#1})}

\newcommand{\sref}[1]{Sec.\,~\ref{#1}}

\newcommand{\cref}[1]{Ref.\,\cite{#1}}

\newcommand{\etal}{{\it et al.}\! }
\newcommand{\apriori}{{\it a priori} }

\newcommand{\Lc}{\mathcal{L}}

\newcommand{\Cbb}{\mathbb{C}}
\newcommand{\Ibb}{\mathbb{I}}
\newcommand{\Jbb}{\mathbb{J}}

\newcommand{\sigmab}{\boldsymbol{\sigma}}

\newcommand{\eb}{\mathbf{e}}

\newcommand{\Bb}{\mathbf{B}}
\newcommand{\Cb}{\mathbf{C}}

\newcommand{\Fb}{\mathbf{F}}

\newcommand{\Mb}{\mathbf{M}}

\newcommand{\Eb}{\mathbf{E}}

\newcommand{\Sb}{\mathbf{S}}
\newcommand{\Ib}{\mathbf{I}}

\newcommand{\tr}{\operatorname{tr}}

\newcommand{\partialb}{\boldsymbol{\partial}}

\newcommand{\defgrad}{\mathbf{F}}

\newcommand{\abstemperature}{T} 

\newcommand{\internalenergy}{\epsilon}

\newcommand{\entropy}{\eta}
\newcommand{\freeenergy}{\Psi}

\title{Polyconvex neural network models of thermoelasticity}
\author{Jan N. Fuhg\\ 
The University of Texas at Austin \\
Austin TX, USA
\And
Asghar Jadoon\\
The University of Texas at Austin \\
Austin TX, USA
\And
Oliver Weeger\\ 
Technical University of Darmstadt \\
Darmstadt, Germany\\
\And
D. Thomas Seidl\\ 
Sandia National Laboratories \\
Albuquerque NM, USA
\And
Reese E. Jones\\ 
Sandia National Laboratories \\
Livermore CA, USA
}

\date{}

\begin{document}

\maketitle

\begin{abstract}
Machine-learning function representations such as neural networks have proven to be excellent constructs for constitutive modeling due to their flexibility to represent highly nonlinear data and their ability to incorporate constitutive constraints, which also allows them to generalize well to unseen data.
In this work, we extend a polyconvex hyperelastic neural network framework to thermo-hyperelasticity by specifying the thermodynamic and material theoretic requirements for an expansion of the Helmholtz free energy expressed in terms of deformation invariants and temperature.
Different formulations which \textit{a priori} ensure polyconvexity with respect to deformation and concavity with respect to temperature are proposed and discussed.
The physics-augmented neural networks are furthermore calibrated with a recently proposed sparsification algorithm that not only aims to fit the training data but also penalizes the number of active parameters, which prevents overfitting in the low data regime and promotes generalization.
The performance of the proposed framework is demonstrated on synthetic data, which illustrate the expected thermomechanical phenomena,  and existing temperature-dependent uniaxial tension and tension-torsion experimental datasets.

\end{abstract}

\section{Introduction}

Many machine learning models have been recently developed for hyperelastic materials, e.g., \cite{klein2022polyconvex,chen2022polyconvex,tac2022data,fuhg2024stress}, while there have been relatively few derived for thermo-hyperelasticity \cite{zlatic2023incompressible} and those typically are limited to a simple temperature dependence; yet, thermo-hyperelasticity is the basis for the thermomechanical response of solid materials.
This gap is, in part, due to the need to develop a general thermo-hyperelastic representation that satisfies thermomechanical principles and constraints.
Generally speaking, embedding physical constraints guides training and provides confidence in the predictions of scientific machine learning (SciML) models.
Neural networks (NNs) are particularly flexible in this regard \cite{fuhg2024stress} unlike, for example, Gaussian Processes \cite{swiler2020survey} or multivariate splines \cite{YVONNET20092723}.

This work is the first, to the authors' knowledge, to embed the fundamental thermodynamic and material-theoretic constraints into a general, complete representation of thermo-hyperelasticity.
The neural network framework is based on a (Helmholtz) free energy that is a function of invariants and satisfies polyconvexity requirements.
It also has the generality to represent temperature-dependent stresses that are not self-similar, as well as thermal expansion, thermal softening and other expected behaviors of a thermoelastic material.
The imposed convexity also promotes smooth derivatives which lead directly to well-behaved stresses.

In the next section, we survey related work.
Then, in \sref{sec:background}, we give a brief exposition of thermomechanics for thermoelastic materials including the physical constraints on the free energy.
Then in \sref{sec:representation} we develop potential free energy representations and describe the proposed physics-constrained neural network representation.
Demonstrations of the efficacy of the proposed framework are given in \sref{sec:demo}, which includes training to complex synthetic data and experimental datasets from the literature.
Finally, we conclude with a summary and avenues for future work in \sref{sec:conclusion}.

\section{Related work}

Classical thermo-hyperelastic models have had a long but sparse history of development.
Early theoretical considerations were postulated by Coleman and Noll \cite{coleman1963thermodynamics},
Truesdell and Noll \cite{truesdell2004non},
and Rivlin \cite{rivlin1986reflections}.
Later, Lu and Pister \cite{lu1975decomposition} made a seminal contribution by introducing the thermal-mechanical multiplicative split \cite{lee1969elastic} of the deformation, which clarifies the thermal effects on the elastic reference configuration.
Lubarda and coworkers \cite{vujovsevic2002finite,lubarda2004constitutive} have postulated complete theories of thermoelasticity based on the Lu and Pister multiplicative decomposition.
Recently, Bouteiller \cite{bouteiller2023complete} developed a general thermoelastic model based on a spherical-deviatoric split and prescribed specific experiments to fully specify its form.
Franke \etal \cite{franke2018energy} and Bonet \etal \cite{bonet2021first} developed specific well-posed polyconvex free energy representations in the context of numerical studies.
Other numerical studies also present thermomechanical models, such as the work of Armero and Simo \cite{armero1992new},  Holzapfel and Simo \cite{holzapfel1996entropy}, and  Tamma and Namburu \cite{tamma1997computational}.
Casey and co-workers have been active in this field \cite{casey1998characterization,casey1998elastic,casey2011nonlinear} making contributions that illuminate a route to construct from measurable quantities the entropy that makes up part of the free energy governing stress.
The entropic contributions of elasticity have been the subject of works by Miehe \cite{miehe1995entropic} and Holzapfel and Simo \cite{holzapfel1996entropy}, while Horgan and Saccomandi \cite{horgan2003finite} proposed a thermo-hyperelastic model based on micromechanical considerations of the extensibility of the underlying polymeric chains.

Micromechanical and statistical models are particularly useful since they give first-principles prescriptions for entropy, which is not easily measurable at a macroscale.
Notable examples of these types of models include: worm-like chains \cite{doi1988theory} and other models describing polymer networks \cite{treloar1975physics,heinrich2003thermoelasticity}, as well as quasiharmonic models \cite{kimmer2007continuum} based on anharmonic phonon contributions.
These models typically require approximations and/or simplified mechanisms to be tractable.

To the authors' knowledge, the only published work that treats thermoelasticity in generality using machine learning is that of
Zlati{\'c} and {\v{C}}anadija \cite{zlatic2023incompressible}.
In their work, they prescribed the form of the temperature dependence and only learned the deformation dependence of an incompressible material with a feedforward multilayer neural network with no physical constraints.
The work of Linka and Kuhl \cite{linka2023new} touches on thermodynamic restrictions to hyperelasticity but does not treat thermal effects such as softening and expansion explicitly.

To construct our representation we rely on neural network architectures that possess both input convexity and input concavity in certain arguments, as well as other fundamental properties such as monotonicity and positivity.
The \emph{Input Convex Neural Network} (ICNN), as well as partially convex and monotonic architectures, were originally proposed by Amos \etal \cite{amos2017input,richter2021input}.
ICNNs have since found applications in measure transport \cite{makkuva2020optimal} and control theory \cite{chen2018optimal}.
In mechanics, ICNNs have seen wide adoption since they aid in learning well-behaved potentials \cite{tac2022data,chen2022polyconvex,as2022mechanics,xu2021learning,klein2022polyconvex,klein2023parametrized,kalina2024neural,fuhg2022learning,fuhg2022machine}.
The work of Linden \etal  \cite{linden2023neural}, discusses the incorporation of physical constraints in neural network-based material models including thermodynamic consistency, balance of angular momentum, objectivity, material symmetry, normalization conditions, coercivity, and ellipticity.
The constraints relevant to the present work will be discussed in the following sections.

\section{Thermoelasticity} \label{sec:background}
A number of general conditions for ensuring a well-behaved material model have been postulated \cite{silhavy2013mechanics}
To ensure ellipticity, polyconvexity is the most well-accepted condition, whereby the internal energy $\hat{\internalenergy}$
\begin{equation}
\internalenergy(\defgrad) = \hat{\internalenergy}(\defgrad, \defgrad^*, \det \defgrad)
\end{equation}
must be convex in each of its arguments, namely the deformation gradient $\defgrad$, its adjugate (matrix of cofactors) $\defgrad^*\equiv \det(\defgrad) \defgrad^{-T}$, and its determinant $\det \defgrad$, which govern the deformation of line segments, areas and volumes, respectively.

For a thermoelastic material, the internal energy depends also on the entropy $\entropy$, and for stable material behavior it is required to be polyconvex in these three deformation inputs and $\entropy$ \cite{silhavy2013mechanics}
\begin{equation}
\internalenergy(\defgrad,\entropy) = \hat{\internalenergy}(\defgrad, \defgrad^*, \det \defgrad, \entropy) \ .
\end{equation}
If the response function is smooth enough, the convexity requirement can be translated into the requirement that the Hessian of $\hat{\internalenergy}$  with respect to each of its arguments is positive semi-definite.

Since entropy $\entropy$ is typically not directly observable, but (the heating measure) absolute temperature $\abstemperature$ is, it is more convenient to work with the Helmholtz free energy $\freeenergy$, which  follows from the Legendre transform of the internal energy
\begin{equation}
\freeenergy(\defgrad,\abstemperature) = \internalenergy(\defgrad,\entropy) - \abstemperature \entropy \ ,
\end{equation}
where $\abstemperature = \partial_\entropy \internalenergy$ and $\entropy = -\partial_\abstemperature \freeenergy$.
This transform implies that the free energy $\hat{\freeenergy}$
\begin{equation}
\freeenergy(\defgrad,\abstemperature) = \hat{\freeenergy}(\defgrad, \defgrad^*, \det \defgrad, \abstemperature)
\end{equation}
is convex in the deformation measures and \emph{concave} in $\abstemperature$.

Objectivity requires $\defgrad$ to be replaced an objective strain measure \cite{silhavy2013mechanics} to eliminate the dependence on the local rotation.
We chose the right Cauchy–Green deformation tensor $\Cb \equiv \defgrad^T \defgrad$ and also assume isotropic material behavior so that the formulation of the free energy can be reduced to dependence on three scalar deformation invariants and temperature:
\begin{equation}
\freeenergy(\defgrad,\abstemperature) = \check{\freeenergy}(I_1,I_2,I_3,\abstemperature) \ ,
\end{equation}
where, again, we have a choice in what invariants to use \cite{fuhg2024stress}.
The Cayley-Hamilton invariants
\begin{equation}
I_1 = \tr \Cb, \quad
I_2 = \tr \Cb^* = \tfrac{1}{2} ( \tr^2 \Cb - \tr \Cb^2) , \quad
I_3 = \det \Cb
\end{equation}
are sufficient, hence the function $\check{\freeenergy}(I_1,I_2,I_3,\abstemperature)$ is required to be
\begin{equation} \label{eq:req}
\begin{matrix*}[l]
\bullet & \text{convex and monotonically increasing in} \ I_{1} \ \text{and} \ I_{2} ,\\
\bullet & \text{convex in} \ I_3 >0 \ \text{or} \  J = \sqrt{I_3} ,\\
\bullet & \text{concave in } \ \abstemperature.
\end{matrix*}
\end{equation}
We will call a free energy that satisfies these conditions polyconvex-concave (PCC).

With this formulation, the second Piola-Kirchhoff $\Sb$ stress can be obtained from
\begin{eqnarray} \label{eq:stress}
\Sb = 2 \partialb_\Cb \freeenergy
&=& 2 \sum_{i=1}^3 \partial_{I_i} \freeenergy \, \partialb_\Cb I_i  \\
&=& 2 \partial_{I_1} \freeenergy \, \Ib
+ 2 \det(\Cb) \left(
\partial_{I_2} \freeenergy \, \left(\tr(\Cb^{-1}) \Cb^{-1}-\Cb^{-2} \right)
+  \partial_{I_3} \freeenergy \, \Cb^{-1}
\right) \nonumber \\
&=& 2 \left( (\partial_{I_1} \freeenergy + I_1 \partial_{I_2} \freeenergy) \, \Ib
- \partial_{I_2} \freeenergy \, \Cb
+  I_3 \partial_{I_3} \freeenergy \, \Cb^{-1}  \right)
\ .
\nonumber
\end{eqnarray}
The corresponding Cauchy stress $\sigmab$ is
\begin{equation} \label{eq:cauchy_stress}
\sigmab = J^{-1} \defgrad \Sb \defgrad^T =
2 \left(J \partial_{I_3} \freeenergy \, \Ib
+ \frac{1}{J} \left( (\partial_{I_1} \freeenergy + I_1 \partial_{I_2} \freeenergy) \, \Bb
- \partial_{I_2} \freeenergy \, \Bb^2 \right) \right) \ ,
\end{equation}
where $\Bb=\defgrad \defgrad^T$ is the left Cauchy-Green deformation tensor.

Some phenomenological expectations are also worth considering.
First, most solid materials expand with increasing temperature, i.e. the volume change at zero pressure is an increasing function of temperature:
\begin{equation}
\partial_\abstemperature J_{p=0} > 0
\ \text{where} \
J_{p=0} = J  \ | \ p(J,\abstemperature) = 0 \ .
\end{equation}
Second, many solid materials soften with increasing temperature i.e. their elastic moduli are decreasing functions of temperature, e.g.
\begin{equation}
\partial_\abstemperature \kappa < 0 \ ,
\end{equation}
where the Lam\'{e} moduli and bulk modulus are defined by
\begin{eqnarray}
\lambda &=& \frac{1}{30} \left(4 \Cbb : \Ibb  - \Cbb : \Jbb \right) \ ,\\
\mu &=& \frac{1}{60} \left( 3 \Cbb:\Jbb - 2 \Cbb : \Ibb \right) \ ,\\
\kappa &=& \frac{1}{9} \Cbb : \Ibb \ ,
\end{eqnarray}
and the elastic tensor is $\Cbb = 4\, \partialb^2_\Cb \Psi$ evaluated at the reference configuration ($\Cb=\Ib$, $\abstemperature=0$).
The components of the isotropic tensors are $\Ibb_{ijkl} = \delta_{ij} \delta_{kl}$ and $\Jbb = \delta_{ik}\delta_{jl} +  \delta_{il}\delta_{jk}$.
Third, the positive heat capacity $c_V \equiv - \abstemperature \, \partial_\abstemperature^2 \freeenergy$ is expected to increase (via the Debye model \cite{ashcroft1976solid}) or remain constant (Dulong-Petit model for insulators \cite{ashcroft1976solid}) with increasing temperature,
which also puts constraints on the form of the free energy.

Many thermoelastic formulations are  predicated on the existence of a multiplicative split of deformation \cite{lu1975decomposition}:
\begin{equation}\label{eq:pisterSplit}
\defgrad = \defgrad_e \defgrad_\abstemperature
=  \defgrad_e \, \underbrace{J_\abstemperature(\abstemperature) \Ib}_{\defgrad_\abstemperature}
= J_\abstemperature(\abstemperature) \defgrad_e \ ,
\end{equation}
which reduces to the last two equalities for an isotropic material.
The thermal deformation $\defgrad_\abstemperature$ is effectively a change of the zero-stress reference configuration from a stress-free ground state $\defgrad = \Ib$, $\abstemperature=0$, $\Sb=\mathbf{0}$.
Hence, scaled invariants from  $\bar{\Cb} = J^{-2/3} \Cb$ that decouple the two deformations are commonly used:
\begin{eqnarray}
J &=& \sqrt{\det\Cb} = J_\abstemperature J_e \ , \\
\bar{I}_1 &=& J^{-2/3}  \tr \Cb \ ,\\
\bar{I}_2 &=& J^{-4/3}  \tr \Cb^* \ .
\end{eqnarray}
For the incompressible, isotropic case, \eref{eq:pisterSplit} is essentially a spherical(thermal)-deviatoric(mechanical) split of the deformation gradient such that $\det\defgrad \equiv J_\abstemperature$.
For the geometric constraint of incompressibility $J_e =1$, the pressure $p = 1/3 \tr \sigmab$ is indeterminant, so all volumetric changes can be attributed to $\defgrad_\abstemperature$.
Furthermore, if the free energy is a homogeneous function of temperature  then
\begin{equation}
\freeenergy(\defgrad=J_\abstemperature \defgrad_e) = \phi(J_\abstemperature) \,\psi_M(\defgrad_e) \ ,
\end{equation}
with a purely temperature-dependent potential $\phi$, and $\psi_M$, which depends only on the elastic deformation.
However, this form is constrained to have \emph{self-similar} mechanical response with temperature, i.e.:
\begin{equation}
\Sb = f(J_\abstemperature(\abstemperature))  (2 \partialb_\Cb \psi_M (\Cb)) \,
\end{equation}
i.e. the stress has a separable dependence on $\abstemperature$ and $\Cb$.

\section{Representation of the free energy} \label{sec:representation}

There are a number of somewhat general representations of thermo-hyperelastic free energy functions in the literature
\cite{miehe1995entropic,franke2018energy,bonet2021first,linka2023new} that use the addition of three terms: (a) a temperature-independent/mechanical-only potential $\freeenergy_M$, (b) a deformation-independent/temperature-only potential $\freeenergy_T$, and (c) a thermomechanical coupling between temperature $\abstemperature$ and volumetric deformation $J$:
\begin{equation}
\freeenergy = \Psi_{M}(I_{1}, I_{2}, J)
+ \phi_\abstemperature(\abstemperature) \psi_M(I_{1}, I_{2}, J)
+ \Phi_{\abstemperature}(\abstemperature)  \ .
\end{equation}
Furthermore, many assume a linear dependence on $\abstemperature$ in the coupling term $\phi_T = -\abstemperature$, ostensibly to satisfy the convexity/concavity requirements and to reproduce the basic thermomechanical phenomenology.

A general complete representation can be found in the tensor product of functions of temperature and deformation
\begin{equation}
\Psi = \sum_{i,j} \Phi_i(\abstemperature) \Psi_j(I_1,I_2,I_3) \ ,
\end{equation}
where $\Phi_i$ and $\Psi_j$ are the basis for well-behaved (square-integrable and separable) Hilbert spaces over $\abstemperature$ and the deformation invariants, respectively.
This is essentially a Schmidt decomposition \cite{kolmogorov1957elements}, which will aid the imposition of the convexity-concavity conditions.
Furthermore, the functional singular value decomposition \cite{bigoni2016spectral,griebel2023analysis} allows for a representation in terms of a diagonalized tensor product of functions, hence we can assume that the finite convex sum
\begin{equation}
\Psi = \sum_{i} \phi_i(\abstemperature) \psi_i(I_1,I_2,I_3)
\end{equation}
has the power to represent all free energy potentials to any selected accuracy.
Here, $\phi_i > 0$ and $\psi_i$ are composed of scaled sums of the basis elements $\Phi_i$ and $\Psi_i$, respectively.
Without loss of generality, we choose to fulfill the requirements in \eref{eq:req} by assuming a form
\begin{equation}\label{eq:Psi_final}
\freeenergy(I_{1}, I_{2}, J, \abstemperature)
= \Psi_{0}(I_{1}, I_{2}, J)
+ \sum_{i=1}^{N_c} \phi_{i}(\abstemperature) \psi_{i}(I_{1}, I_{2}, J)
+  \Phi_{T}(\abstemperature)  \ ,
\end{equation}
which is polyconvex-concave (PCC)if
\begin{itemize}
\item $\freeenergy_{0}$ and $\psi_{i}$ are positive and convex, non-decreasing in $I_{1}$ and $I_{2}$ and convex in $J$,
\item $\phi_{i}$ and $\Phi_{T}$ are concave and positive.
\end{itemize}
Note, the first term $\freeenergy_{0}$ can be seen as part of the sum where $\phi_0 \equiv 1$.
The function $\Phi_{T}$ will be undetermined by stress data, but it plays a role in determining the heat capacity and the overall concavity of $\freeenergy$ with respect to temperature.
Also, if we constrain $\phi_i(0)\equiv 0$, then $\freeenergy_0$ can be interpreted as the ground state elastic potential; however, this would hamper the discovery of a separable free energy function as in \sref{sec:treloar}.
We furthermore remark that if the second derivative of $\phi_{i}(T)$ is zero, then $\Phi_{T}(T)$ is no longer required to be positive.
The number of terms in the expansion in \eref{eq:Psi_final} is a hyperparameter of the material model that can be determined through regularization or a greedy algorithm with cross-validation.
Also, we impose the smoothness requirements $\psi_i \in C^1$ such that a well-defined stress $\Sb = \partial_\Cb \freeenergy$ can be obtained from $\freeenergy$.

We aim to describe all the functions in \eref{eq:Psi_final} by physics-augmented neural networks.
The general form of a feedforward neural network with $L-1$ hidden layers, input $\bm{x}_{0} \in \mathbb{R}^{n^{0}}$ and scalar output $x_{L}\in \mathbb{R}$ is
\begin{equation}
\begin{aligned}
\bm{x}_{0} &\in \mathbb{R}^{n^{0}} \ ,\\
\bm{x}_{1} = \mathcal{A}_{1} \left( \bm{x}_{0} \bm{W}_{1}^{T} + \bm{b}_{1} \right) &\in \mathbb{R}^{n^{1}} \ ,\\
\bm{x}_{l} = \mathcal{A}_{l} \left( \bm{x}_{l-1} \bm{W}_{l}^{T} + \bm{b}_{l} \right) &\in \mathbb{R}^{n^{l}} \ , \qquad l=2, \ldots, L-1 \ , \\
x_{L} = \bm{x}_{L-1} \bm{W}_{L}^{T} + \bm{b}_{L}, &\in \mathbb{R} \ .
\end{aligned}
\end{equation}
Here, we have defined the network activation functions as $\mathcal{A}_{l}:\mathbb{R}^{n^{l}}\to\mathbb{R}^{n^{l}}$, the weights as $\bm{W}_{l}\in \mathbb{R}^{n^{l}\times n^{l-1}}$, and the biases as $\bm{b}_{l}\in \mathbb{R}^{n^{l}}$.
We can constrain this model form to derive functions with our required properties.
By assuming a reparametrization of the trainable parameters using a smoothed \emph{gating} system (c.f. Louizos \etal \cite{louizos2017learning} and Fuhg \etal \cite{fuhg2023extreme}) the number of non-zero weights and biases can be pruned using a form of smoothed $L^{0}$ regularization. To regularize our networks and prevent overfitting, we have used this concept in the following when only limited experimental data in specific stress states is available.

\paragraph{Network construction for $\freeenergy_{0}$ and $\psi_{i}$.}
Following \cref{amos2017input}, the output of the network $x_L$ is convex with regards to its input $x_0$ if the weights $\bm{W}_{l}$ with $l=2, \ldots ,L$ are non-negative and all the activation functions are convex and non-decreasing.
In this work, we use the Softplus activation function $\mathcal{A}_{l}(x) =\log (1 + \exp x)$ for all the layers (in a component-wise fashion).
Additionally, due to the positivity of the Softplus function, if the bias vector of the output is positive $ \bm{b}_{L} \geq 0$, then the network output is positive.
Lastly, the network output is non-decreasing if $\bm{W}_{1} \geq0$ is also enforced \cite{klein2022polyconvex}.
Following Linden \etal \cite{linden2023neural},  in order to be able to be able to represent negative stresses, the additional invariant $-2J$ is used as an input to both networks.

\paragraph{Network construction for $\phi_{i}$.}
We consider two different approaches to fulfill the requirements for polyconvexity-concavity (PCC):
\begin{enumerate}
\item As discussed before, one option is to constrain the networks $\phi_{i}$ to be positive and concave.
The first derivative of a scalar network output $x_{L} \in \mathbb{R}$ with respect to a scalar positive input $x_{0} \in \mathbb{R}_{\geq 0}$ is given by \cite{ratku2022derivatives,fuhg2023modular}:
\begin{equation}
\frac{d x_{L}}{d x_{0}} = \prod_{l=0}^{L-1} \left[ ({\mathcal{A}}_{L-l}^{'}( \underbrace{\bm{x}_{L-l-1} \bm{W}_{L-l}^{T} + \bm{b}_{L-l}}_{\bm{y}_{L-l}}))^{T} \bm{j}_{L-l} \right] \circ \bm{W}_{L-l}  \ ,
\end{equation}
where $\bm{j}_{l}$ are row vectors of ones with same size as $\bm{x}_{l-1}$.
The second derivative of the network output is given by:
\begin{equation}
\frac{d^{2} x_{L}}{d^{2} x_{0}} = \sum_{l=1}^{L} \bm{J}N^{l+1,L} \left(   \left[ \left\{ (\mathcal{A}_{l}^{''})^{T} \circ \bm{m} \right\}  \bm{j}_{l} \right] \circ \bm{W}_{l} \right) \bm{J}N^{1,l-1} \ ,
\end{equation}
where
\begin{equation}
\begin{aligned}
\bm{J}N^{p,q} &=  \prod_{k=L-q}^{L-p} \left[ \Big(\mathcal{A}_{L-k}^{'}( \bm{x}_{L-k-1} \bm{W}_{L-k}^{T} + \bm{b}_{L-k}) \Big)^{T} \bm{j}_{L-k} \right] \circ \bm{W}_{L-k} \ ,  \\
\mathcal{A}_{l}^{''} &= \mathcal{A}_{l}^{''}( \bm{x}_{l-1} \bm{W}_{l}^{T} + \bm{b}_{l}) \ , \\
\bm{m} &= \bm{W}_{l} \bm{J}N^{1,l-1} \ .
\end{aligned}
\end{equation}
In order for the networks to be positive and concave we require that $x_{L}>0$ and $\frac{d^{2} x_{L}}{d^{2} x_{0}}\leq 0$.
We can achieve this by enforcing the constraints:
\begin{itemize}
\item $\bm{W}_{l} \geq 0$ and $\bm{b}_{l}\geq 0$ for $l=1, \ldots L$ ,
\item $\mathcal{A}_{l}: \mathbb{R}_{+} \rightarrow \mathbb{R}_{+}$ for $l=1, \ldots L-1$ ,
\item $\mathcal{A}_{l}': \mathbb{R}_{+} \rightarrow \mathbb{R}_{+}$ for $l=1, \ldots L-1$ ,
\item $\mathcal{A}_{l}'': \mathbb{R}_{+} \rightarrow \mathbb{R}_{-}$ for $l=1, \ldots L-1$ .
\end{itemize}
A possible option for the activation function is the logistic function
\begin{equation}
\mathcal{A}(x) = \frac{1}{1+\exp(-x)} \ ,
\end{equation}
which is positive and has the derivatives:
\begin{eqnarray}
\mathcal{A}'(x) &=& \frac{\exp(-x)}{\left(1+\exp(-x) \right)^{2}} \ , \\
\mathcal{A}''(x) &=& \frac{2 \exp(-2x)}{\left(\exp(-x)+1 \right)^{3}} - \frac{\exp(-x)}{\left(1+\exp(-x) \right)^{2}} \ .
\end{eqnarray}

One problematic aspect of these constraints is that the network is overconstrained to be positive, non-decreasing, and concave, which appears to be only possible over a finite range or by a constant function.
However, since the reciprocal of a positive, non-decreasing, and concave function is positive, non-increasing, and concave we can model decreasing functional forms of $\phi_{i}$ by taking the reciprocal of the network output.
This would require \apriori knowledge of the functional form of $\phi_{i}$.

\item A second option is to enforce the functions $\phi_{i}$ to be positive and $\partial^2_\abstemperature \phi_{i}=0$.
This can, for example, be achieved by choosing a piecewise linear functional form that guarantees positivity.
We can construct a neural network that has similar characteristics by ensuring that the weight and bias of the output layer are positive $\bm{W}_{L}\geq 0$ and $\bm{b}_{L}\geq 0$, by choosing a positive activation function in the last hidden layer $\mathcal{A}_{L-1}\geq 0$, and by guaranteeing that the second derivative of all activation functions is zero, i.e. $\mathcal{A}_{l}''=0$.
Possible options for $\mathcal{A}_{L-1}$ are
\begin{equation}
\begin{aligned}
\mathcal{A}_{L-1}(x) = \max(0,x), \quad  \mathcal{A}_{L-1}(x) = |x| ,\quad \mathcal{A}_{L-1}(x)=\max(-x,0) \ .
\end{aligned}
\end{equation}
For the remaining activation functions $\mathcal{A}_{k}$ with $k\neq L-1$, we can also include negative linear functions such as
\begin{equation}
\mathcal{A}_{k}(x) = \max (0,x) + a \min (0,x), \quad a\in \mathbb{R} \ .
\end{equation}

In this case, in order to ensure polyconvexity, $\freeenergy_{T}$ needs to be concave in $T$.
In the following, we pursue this second option and employ $ReLU$ activation functions $\mathcal{A}_{l}(x) = \max(0,x)$ throughout.
\end{enumerate}

\paragraph{Network construction for $\freeenergy_{T}$.}
As mentioned, in this work we assume that only stress data over different temperatures is available; hence, $\freeenergy_{T}$ is left undetermined.
However, the function has to be positive and concave if the second derivative of at least one of the networks $\phi_{i}$ is non-zero.
This can be achieved with the network architecture described for $\phi_{i}$. Otherwise a concave $\freeenergy_{T}$ is sufficient for PCC.

\section{Demonstrations} \label{sec:demo}
To demonstrate the efficacy of the proposed machine learning framework for thermo-hyperelastic material modeling, we use two synthetic datasets where all components of strain and temperature are controlled and all stress components are observed, as well as three experimental datasets with limited observations \cite{zhang2018temperature, mohsin1987thermoelastic,fu2021ability}.
Although it is typical to use an empirical temperature scale, we will maintain our use of an absolute temperature $\abstemperature$ and employ $\abstemperature=0$ for the reference configuration.

\subsection{Synthetic data}
For these examples, we use analytic free energy data generating models.
To obtain the training dataset we sample $(\Fb,\abstemperature)$ uniformly as $[\defgrad-\Ib]_{ij} \in \mathcal{U}[0.6,1.4]$ and $\abstemperature \in \mathcal{U}[0,2]$, and calculate the resulting $\Cb$  and $\Sb$.
Here $\mathcal{U}(a,b)$ is a uniform distribution over the range $[a,b]$ and $\abstemperature=1$ is nominally room temperature.
For the training of the models on the generated data sets with $N_D$ data points, we use a mean squared loss on training samples $\Sb_i$ of the stress tensor:
\begin{equation}
\Lc = \frac{1}{N_D} \sum_{i=1}^{N_D} \| \Sb_i - 2 \partialb_\Cb \freeenergy(\Cb_i,T_i)  \|^2 \ .
\end{equation}
Recall that training on stress data implies that we only determine $\freeenergy$ up to a function of temperature $\freeenergy_\abstemperature(\abstemperature)$, c.f.~\eref{eq:Psi_final}.
In the following, the neural network models related to the mechanical components $\Psi_{0}$ and $\psi_{i}$ are composed of 2 hidden layers with 30 neurons with Softplus activation function.
The temperature-dependent networks $\psi_{i}$ consist of 2 layers with 40 neurons with ReLU as the activation function.
We employ the classical ADAM optimizer \cite{kingma2014adam} with a learning rate of $10^{-3}$.

\subsubsection{Neo-Hookean material}
As mentioned in \sref{sec:representation}, some thermo-hyperelastic models in the literature \cite{bonet2021first} have free energies of the form
\begin{equation} \label{eq:bonet_free_energy}
\freeenergy = \freeenergy_0(\Cb) - \abstemperature \entropy(J,\abstemperature) + \Phi_\abstemperature(\abstemperature) \ ,
\end{equation}
where the entropy contribution $-\abstemperature \entropy$ provides coupling between the deformation and temperature.
For this type of decomposition, the second Piola-Kirchhoff stress $\Sb$ is
\begin{equation}
\Sb = 2 \partial_\Cb \freeenergy_0(\Cb) - 2 \abstemperature \partial_J \entropy \partialb_\Cb J
= \underbrace{2 \partial_\Cb \freeenergy_0(\Cb)}_{\Sb_0} -  \abstemperature J \partial_J \entropy \Cb^{-1} \ ,
\end{equation}
and the Cauchy stress $\sigmab$ is
\begin{equation}
\sigmab = \underbrace{2 J^{-1} \defgrad \partial_\Cb \freeenergy_0(\Cb) \defgrad^T}_{\sigmab_0}  + \abstemperature J \partial_J \entropy \Ib \ .
\end{equation}
Note that the entropy contribution is zero at the absolute temperature $\abstemperature=0$ so that $\Sb_0$ and $\sigmab_0$ are the zero temperature stress response derived from $\freeenergy_0$.
Also, if the entropy is not temperature-dependent, the stress response is self-similar with respect to $\abstemperature$.

In particular, we take $\freeenergy_0$ to be the potential for a Rivlin-type \cite{rivlin1997collected} compressible  neo-Hookean model
\begin{equation}
\freeenergy_0 = \frac{1}{2} \mu \bar{I}_1 + \frac{1}{2} \kappa (J-1)^2 \ ,
\end{equation}
where $\bar{I}_1 = J^{-2/3} \tr \Cb = I_3^{-1/3} I_1  > 0$ and $J = \sqrt{\det \Cb} > 0$.
This potential is convex in $I_1$ and $J$.
We take the material properties $\kappa=0.73$ and $\mu=0.41$ to be temperature-independent (normalized) constants representative of a polymer with Poisson's ratio of 0.4.
The entropy contribution
\begin{equation}
\abstemperature \entropy= \abstemperature \left(  c_v \Gamma_0 \frac{J^q-1}{q}  \right)
\end{equation}
follows from the Mie-Gruneisen model \cite{mie1903kinetischen,gruneisen1912theorie}.
We set $c_v \Gamma_0 = 0.1$ and $q = 1.5$.
This free energy satisfies the PCC requirements \eqref{eq:req}.
Finally the temperature-only contribution
\begin{equation} \label{eq:Phi}
\Phi_\abstemperature = c_V (\abstemperature -\abstemperature_0 - \abstemperature \log(\abstemperature/\abstemperature_0))
\end{equation}
is chosen to be consistent with expectations for a \emph{constant} heat capacity $c_V = \abstemperature\, \partial^2_\abstemperature \freeenergy$, which does not affect the stress.

The resulting stress measures are
\begin{equation}
\Sb = \mu J^{-2/3} (\Ib - 1/3 I_1 \Cb^{-1})
+ \left( \kappa (1-J^{-1}) + \abstemperature c_v \Gamma_0 J^{q} \right) \Cb^{-1}
\end{equation}
and
\begin{equation}
\sigmab = \mu J^{-5/3} (\Bb - 1/3 I_1 \Ib) + \left(\kappa (J-1) +  \abstemperature c_v \Gamma_0 J^{q-1} \right) \Ib \ ,
\end{equation}
where the partial derivatives of the potential are:
\begin{eqnarray}
\partial_{I_1} \freeenergy &=&  \mu I_3^{-1/3} \ ,\\
\partial_{I_3} \freeenergy &=& - \frac{2}{3} \mu I_1 I_3^{-4/3} + \kappa \frac{J-1}{J} + c_v \Gamma_0 J^{q-2} \abstemperature \ .
\end{eqnarray}

Thermal expansion is given by the solution of
\begin{equation}
p  =
3 \left( (\kappa (J^2 - J)
- c_v \Gamma_0 J^{q} \abstemperature \right) = 0
\end{equation}
for $J$, given $\abstemperature$, which is the inverse of
\begin{equation} \label{eq:bonet_thermal_expansion}
\abstemperature(J) =
\frac{\kappa}{c_v \Gamma_0}
\frac{J^2-J}{J^q} \ .
\end{equation}
For the special case of $q=3/2$, \eref{eq:bonet_thermal_expansion} has the explicit inverse
\begin{equation} \label{eq:bonet_thermal_expansion2}
J(\abstemperature) = 1 + \frac{1}{2} \left( \left(\frac{c_v \Gamma_0}{\kappa}\right)^2 \abstemperature^2 + \sqrt{
\left(\frac{c_v \Gamma_0}{\kappa}\right)^2 \abstemperature^2 +
\frac{1}{4}\left(\frac{c_v \Gamma_0}{\kappa}\right)^4 \abstemperature^4
}\right) \ .
\end{equation}

\begin{figure}
\centering
\includegraphics[width=\textwidth]{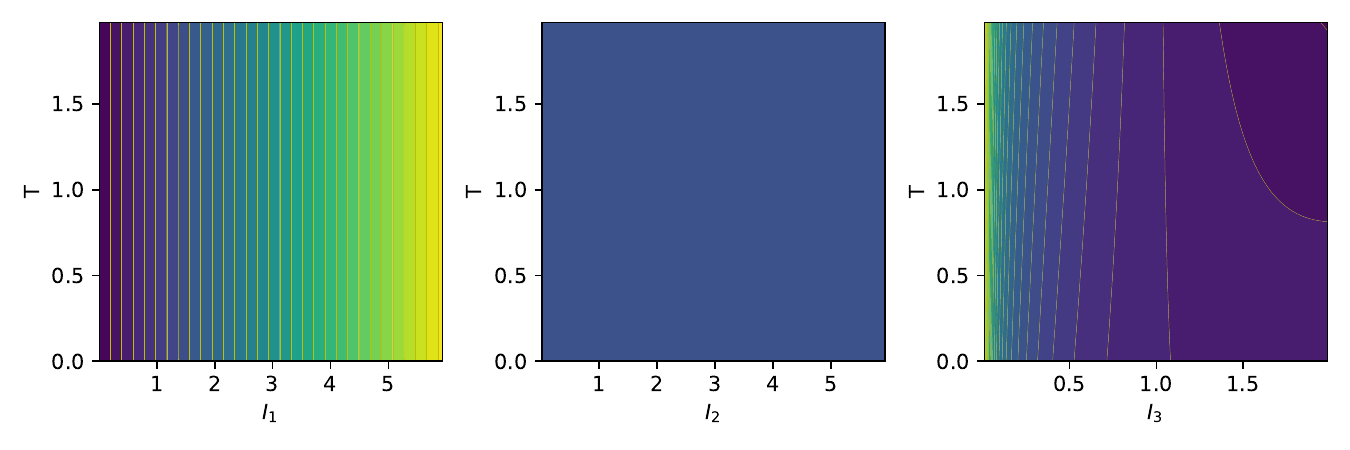}

\caption{Neo-Hookean potential.
Left: $\freeenergy(I_1,3,1,\abstemperature)$.
Middle: $\freeenergy(3,I_2,1,\abstemperature)$.
Right: $\freeenergy(3,3,I_3,\abstemperature)$.
Note the selected potential does not have a $I_2$ dependence.
}
\label{fig:bonet}
\end{figure}

\begin{figure}
\centering
\includegraphics[width=0.99\textwidth]{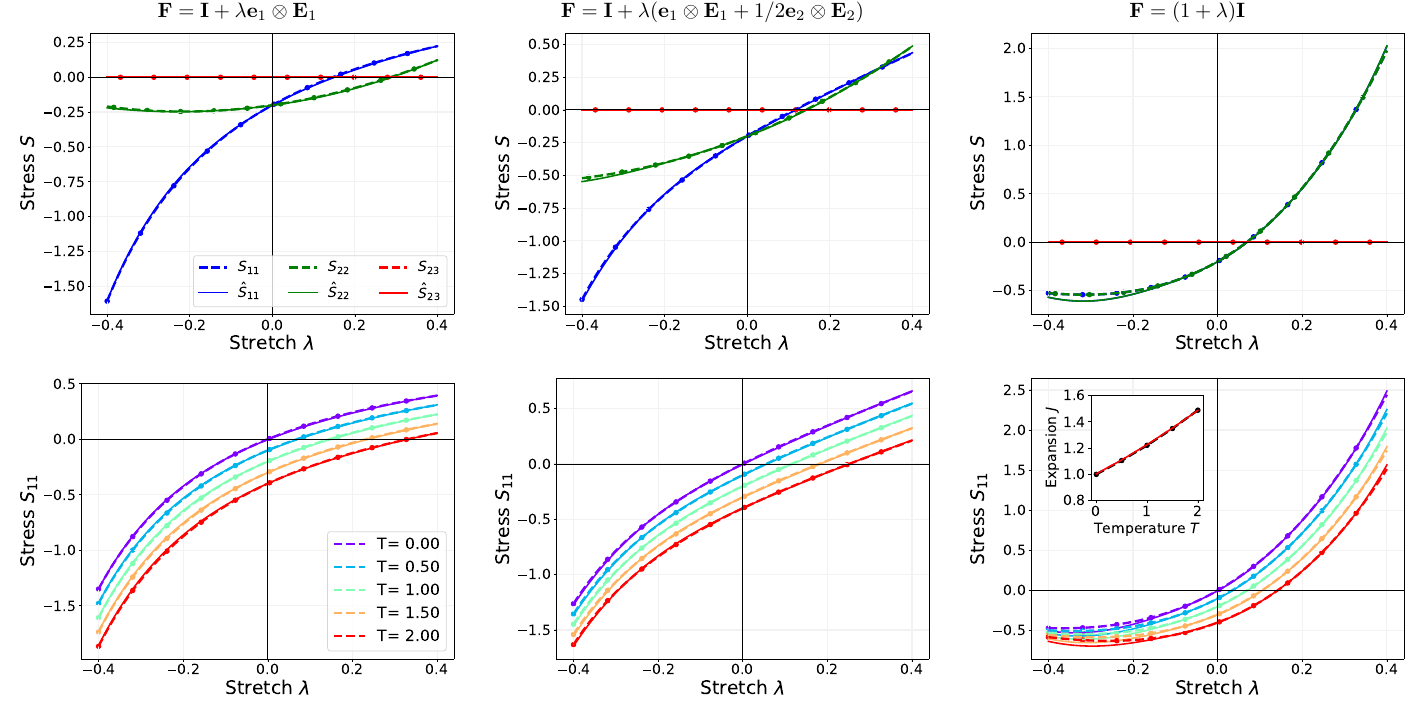}
\caption{Neo-Hookean comparison, dashed-dotted lines: data, solid line: predictions.
Upper: selected components.
Lower: for various temperatures.
Left: uniaxial stretch, middle: biaxial stretch, right: volumetric deformation.
Inset: thermal expansion.
}
\label{fig:bonet_nn}
\end{figure}

Projections of the analytical potential are shown in \fref{fig:bonet}.
The trained model's correspondence with held-out experimentally realizable loading modes: (a) isothermal uniaxial stretch $\Fb = \Ib + \lambda \eb_1 \otimes \Eb_1$, (b) unequal biaxial stretch $\Fb = \Ib + \lambda ( \eb_1 \otimes \Eb_1 + 1/2 \eb_1 \otimes \Eb_1)$, and (c) volumetric stretch $\Fb = (1+\lambda) \Ib$ is shown in \fref{fig:bonet_nn}.
Both the selected stress components at a fixed temperature and the 11-stress component for a sequence of temperatures illustrate the accuracy of the model on this validation data.
Since the stress response was self-similar with temperature, one coupled term $N_c=1$ was sufficient for an accurate representation.

\subsubsection{Saint Venant material}
A St. Venant model \cite{lubarda2004constitutive} has a free energy
\begin{equation} \label{eq:StV}
\freeenergy = \frac{1}{2} \Eb : \Cbb \Eb -\abstemperature \Mb : \Eb + \Phi_\abstemperature(\abstemperature) \ ,
\end{equation}
where $\Eb = 1/2 (\Cb -\Ib)$ is the Lagrange strain tensor and resembles the classical linear thermoelastic model in form.
For isotropic materials, the thermal expansion tensor $\Mb = \alpha \Ib$ is proportional to the identity tensor $\Ib$ and  $\Cbb = \lambda \Ibb + \mu \Jbb$, so that \eref{eq:StV} reduces to
\begin{equation}
\freeenergy = \frac{1}{2} \lambda(\abstemperature) \left( \tr \Eb \right)^2 + \mu(\abstemperature) \tr \Eb^2 -  \frac{1}{2} \underbrace{ \abstemperature \alpha(\abstemperature)}_{\gamma(\abstemperature)} \tr \Cb + \Phi(\abstemperature) \ ,
\end{equation}
where we have absorbed $1/2 \gamma(\abstemperature) \tr \Ib$ into $\Phi_\abstemperature(\abstemperature)$.
The second Piola-Kirchhoff stress is
\begin{equation}
\Sb = (\lambda \tr \Eb - \gamma) \Ib + \mu \Eb \ ,
\end{equation}
which, like the previous model, has thermal expansion effects added to a temperature-independent elastic response.
However, the resulting Cauchy stress
\begin{equation}
\sigmab = \frac{1}{J}\left( (\frac{1}{2}\lambda ( \tr \Bb -3) - \gamma ) \Bb +  \frac{1}{2} \mu (\Bb^2 - \Bb)
\right)
\end{equation}
has the coefficient controlling thermal expansion $\gamma$ paired with the right Cauchy-Green stretch $\Bb = \defgrad \defgrad^T$.
Since the deformation components are positive and convex, the coefficients $\lambda$, $\mu$, and $\kappa$ need to be positive and concave with respect to $\abstemperature$.
Although being convex in $I_1$ and $I_2$ this potential does not satisfy the monotonicity requirements \eqref{eq:req} of a PCC free energy.

For the particular temperature dependence of the moduli we use:
\begin{eqnarray}
\lambda &=& \lambda_0 \frac{ \tanh(\abstemperature_c - \abstemperature) +1}{\tanh(\abstemperature_c ) +1} \ ,\\
\mu &=& \mu_0 \frac{ \tanh(\abstemperature_c - \abstemperature) +1}{\tanh(\abstemperature_c ) +1} \ ,\\
\gamma  &=& \gamma_0  \left(\frac{\abstemperature}{\abstemperature_0}\right)^a \ ,
\end{eqnarray}
with $\lambda_0 = 0.73$, $\mu_0 = 0.41$, $\kappa_0 = \lambda_0 + 2/3 \mu_0 = 1.0$, $\gamma_0=0.2$, $a=1/2$, $\abstemperature_0 = 1.0$, and $\abstemperature_c = 2 \abstemperature_0$.
Note that $\tanh(a-x)$ is concave for $x<a$ so $\lambda$ and $\mu$ are concave in $\abstemperature \in [0, \abstemperature_c]$.
Furthermore, it is $\lambda(0) = \lambda_0$ and $\lim_{\abstemperature\to\infty} \lambda = 0$, and likewise relations hold for $\mu$.
The coefficient controlling the thermal expansions starts with $\gamma(0) = 0$ but grows with temperature.
The different temperature dependencies of $\gamma$ and $\kappa = \lambda + 2/3 \mu$ ensure that the resulting stresses are not self-similar with respect to $\abstemperature$.

Thermal expansion is given by the solution of
\begin{equation}
p =
\frac{3}{2} J \bigl(
(J^{2/3}-1)
\underbrace{\left(3 \lambda(\abstemperature)
+ 2 \mu(\abstemperature)
\right)}_{3 \kappa(\abstemperature)}
- 2 \gamma(\abstemperature)
\bigr) = 0
\end{equation}
for $J$, given $\abstemperature$, which is:
\begin{equation}
J(\abstemperature) = \left(1 + \frac{2 \gamma(T)}{3\kappa(T)}\right)^{3/2}  \ .
\end{equation}

\begin{figure}
\centering
\includegraphics[width=\textwidth]{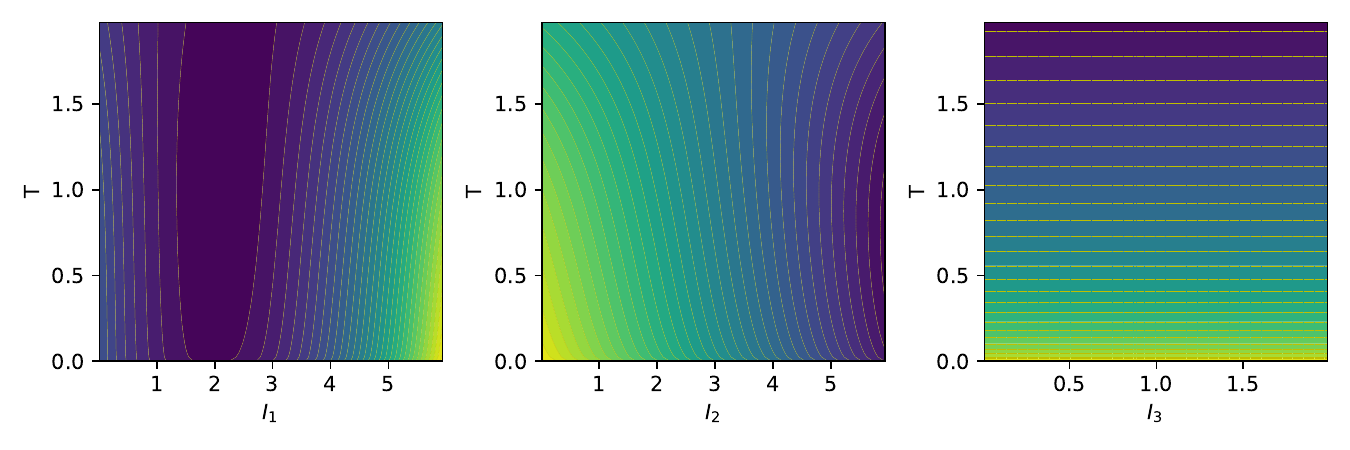}
\caption{Saint Venant potential.
Left: $\freeenergy(I_1,3,1,\abstemperature)$.
Middle: $\freeenergy(3,I_2,1,\abstemperature)$.
Right: $\freeenergy(3,3,I_3,\abstemperature)$.
}
\label{fig:stv}
\end{figure}

\begin{figure}
\centering
\includegraphics[width=0.99\textwidth]{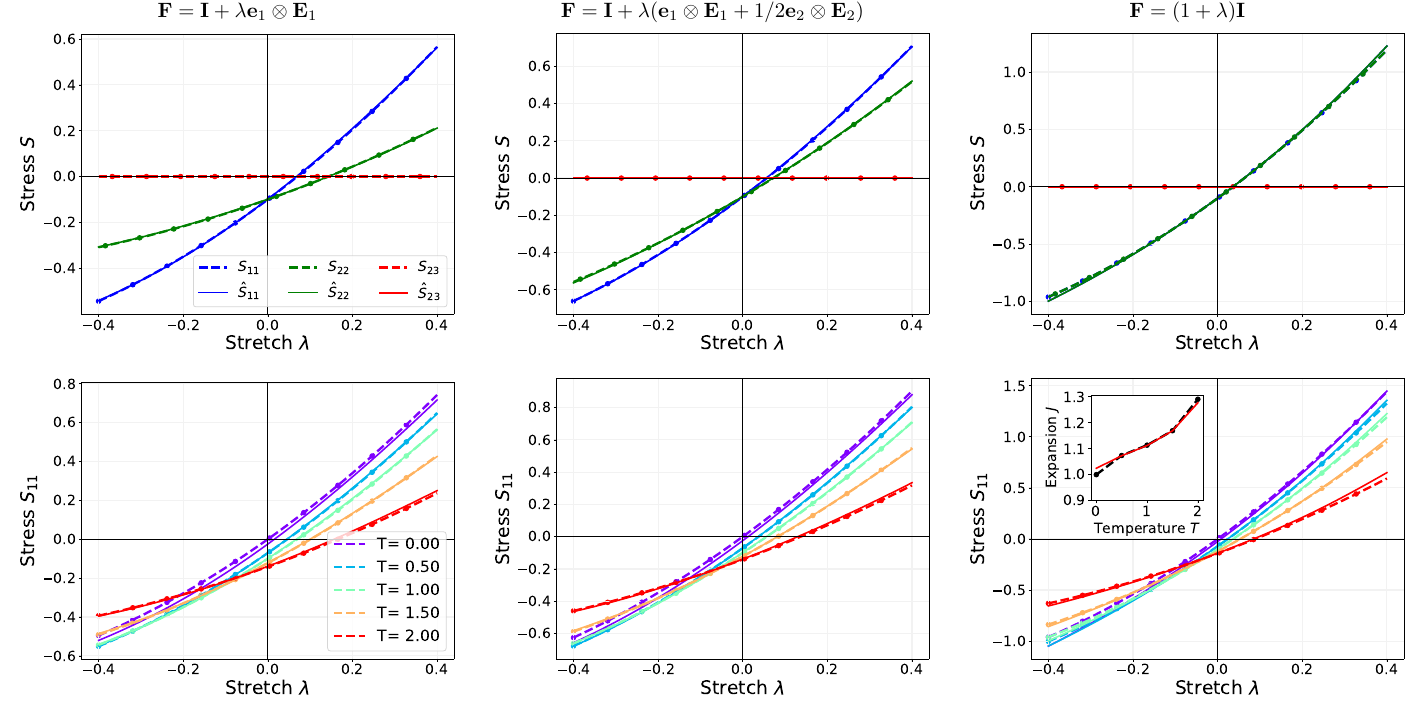}
\caption{Saint Venant comparison,  dashed-dotted lines: data, solid line: predictions. Upper: selected components.
Lower: for various temperatures. Left: uniaxial stretch. Middle: biaxial stretch. Right: volumetric deformation.
}
\label{fig:stv_nn}
\end{figure}

Projections of the analytical potential are shown in \fref{fig:stv}.
We used a greedy algorithm based on cross-validation error to arrive at a two-coupled term $N_c=2$ fit, c.f.~\eref{eq:Psi_final}.
An $L^{p}$ regularization approach might have been equally effective at determining how many coupling terms are needed for a sufficient accurate representation.
The trained model correspondence with held-out validation data is given in \fref{fig:stv_nn}, which shows the ability of the representation to capture thermal softening and thermal expansion trends for a response that does not have a self-similar response with temperature.

\subsection{Experimental data}
To demonstrate the application of the proposed NN framework to modeling experimental data, we selected three datasets from the literature that have limited observations of the stress state of the specimens at various temperatures.
The classic experiment of Moshin and Treloar \cite{mohsin1987thermoelastic} provides data on extension and twist-constrained rubber specimens subject to temperature loading (\sref{sec:treloar}).
Recently
Zhang \etal  \cite{zhang2018temperature} collected tensile data of soft tissue over a range of temperatures (\sref{sec:zhang}).
Lastly, we study a dataset provided by Fu \etal \cite{fu2021ability} measuring the uniaxial tension of carbon black rubber under different temperatures (\sref{sec:fu}).

In the following, we used one mechanical-thermal coupling term in the representation of the $\freeenergy$, i.e. $N_{c}=1$ in \eref{eq:Psi_final}, so that $\freeenergy$ is comprised of three networks: $\Psi_{0}$, $\psi_{1}$ and $\phi_{1}$ which are trained simultaneously.
We remark that we have regularized the neural networks for all of the following examples by pruning the number of trainable parameters with $L^{0}$ regularization to avoid overfitting.
For the experimental data, all network models consist of 2 hidden layers with 20 neurons.
The activation function is chosen as Softplus for $\Psi_{0}$ and $\psi_{1}$ and as ReLU for $\phi_{1}$.
The optimizer and learning rate remain unchanged from the synthetic examples.
The regularization parameter is set to $2 \cdot 10^{-4}$.
\subsubsection{ Tension-torsion of butyl rubber} \label{sec:treloar}
Tension-torsion is a common experimental setup since it can probe both the extensional and shear response of a material at once.
The classic Moshin and Treloar experiment \cite{mohsin1987thermoelastic} provides force and moment versus extension and twist data for a set of isothermal experiments on butyl rubber.

For this demonstration, we employed a mean squared loss on the force $F$ and moment $M$ for the tension-torsion data $\{ \gamma_i, \tau_i, \abstemperature_i, F_i, M_i \}$
\begin{equation}
\Lc = \frac{1}{N_D} \sum_{i=1}^{N_D} \big( | F_i - F(\gamma_i,\tau_i, \abstemperature_i)  |^2 + | M_i - M(\gamma_i,\tau_i, \abstemperature_i) |^2 \big)
\end{equation}
which is amended by an $L^{0}$ regularization term following Fuhg \etal \cite{fuhg2023extreme}.
We only observe the end force $F$ and moment $M$ for a given axial extension $\gamma$ and twist per stretched length $\tau$.
We use an analytical solution \cite{rivlin1948large,rivlin1951large,hartmann2001numerical,kirkinis2002extension} for an incompressible hyperelastic material:
\begin{eqnarray}
F &=& 2 \pi  \int_0^{R_o} \left(
( 1 - \gamma^3 + 1/2 \gamma R^2 \tau^2) \gamma^{-3/2}
\partial_{I_1} \Psi
+ (-1 + \gamma^3 i - \gamma^2 R^2 \tau^2) \gamma^{-5/2}
\partial_{I_2} \Psi \right)
R \, \mathrm{d}R \label{eq:F} \ ,
\\
M &=& 2 \pi \tau \int_0^{R_o} \left(
\gamma^{ 1/2} \partial_{I_1} \Psi
+ \gamma^{-1/2} \partial_{I_2} \Psi
\right) R^3\, \mathrm{d}R \label{eq:M} \ ,
\end{eqnarray}
which is based on the ansatz for the motion:
\begin{eqnarray}
r&=& \gamma^{-1/2} R \ ,\\
\vartheta &=& \Theta + \gamma \tau Z  \ ,\\
z &=& \gamma Z \ ,
\end{eqnarray}
where $R,\Theta,Z$ are referential cylindrical polar coordinates and
\begin{eqnarray}
I_1 &=& 2 \gamma^{-1} + \gamma^2 + \gamma \tau^2 R^2 \ ,\\
I_2 &=& \gamma^{-2} + 2 \gamma +  \tau^2 R^2 \ ,\\
I_3 &=& 1
\end{eqnarray}
are the corresponding invariants.
We use quadrature to evaluate the integrals in \eref{eq:F} and \eref{eq:M} in terms of the NN representation of $\freeenergy$.

\begin{figure}
\centering
\begin{subfigure}{0.5\textwidth}
\centering
\includegraphics[scale=0.35]{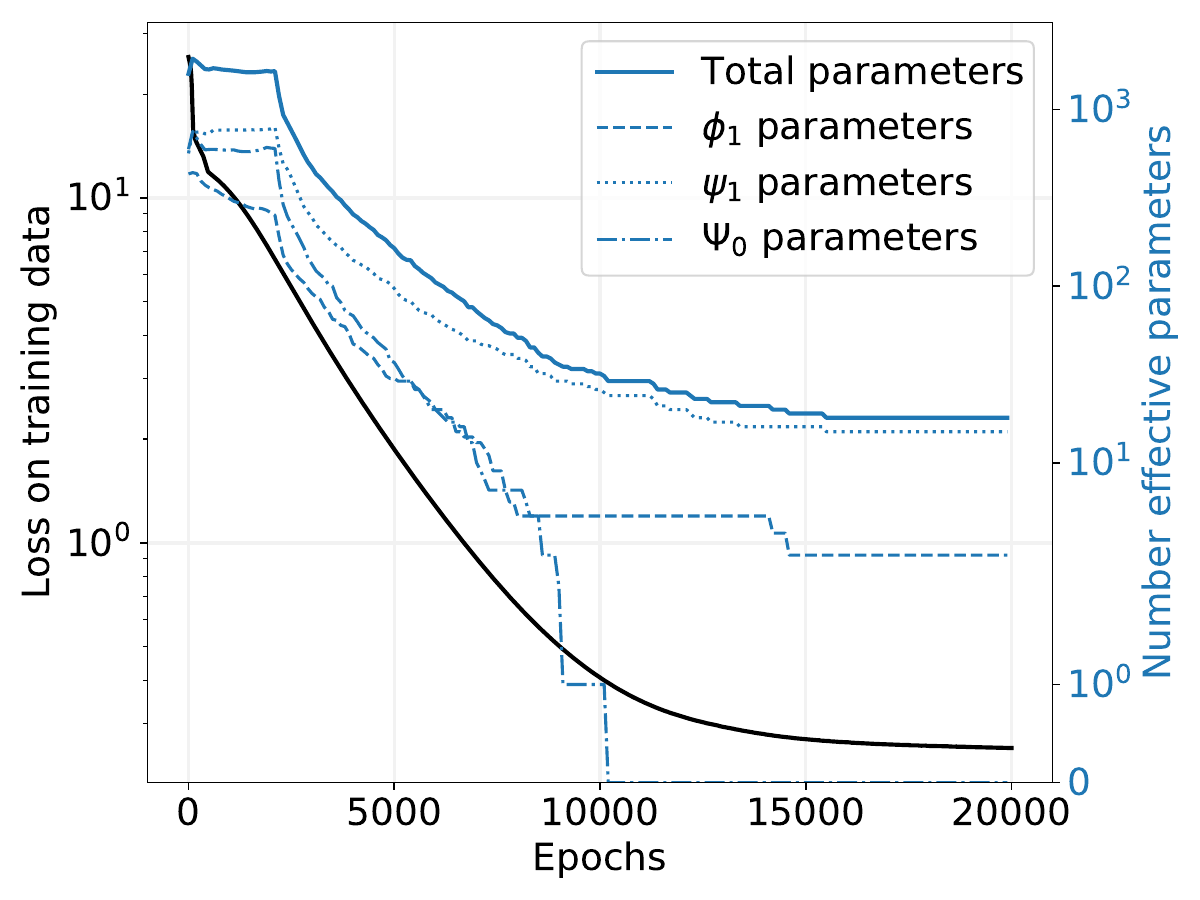}
\caption{}\label{fig:tension-torsion_loss_a}
\end{subfigure}%
\begin{subfigure}{0.5\textwidth}
\centering
\includegraphics[scale=0.35]{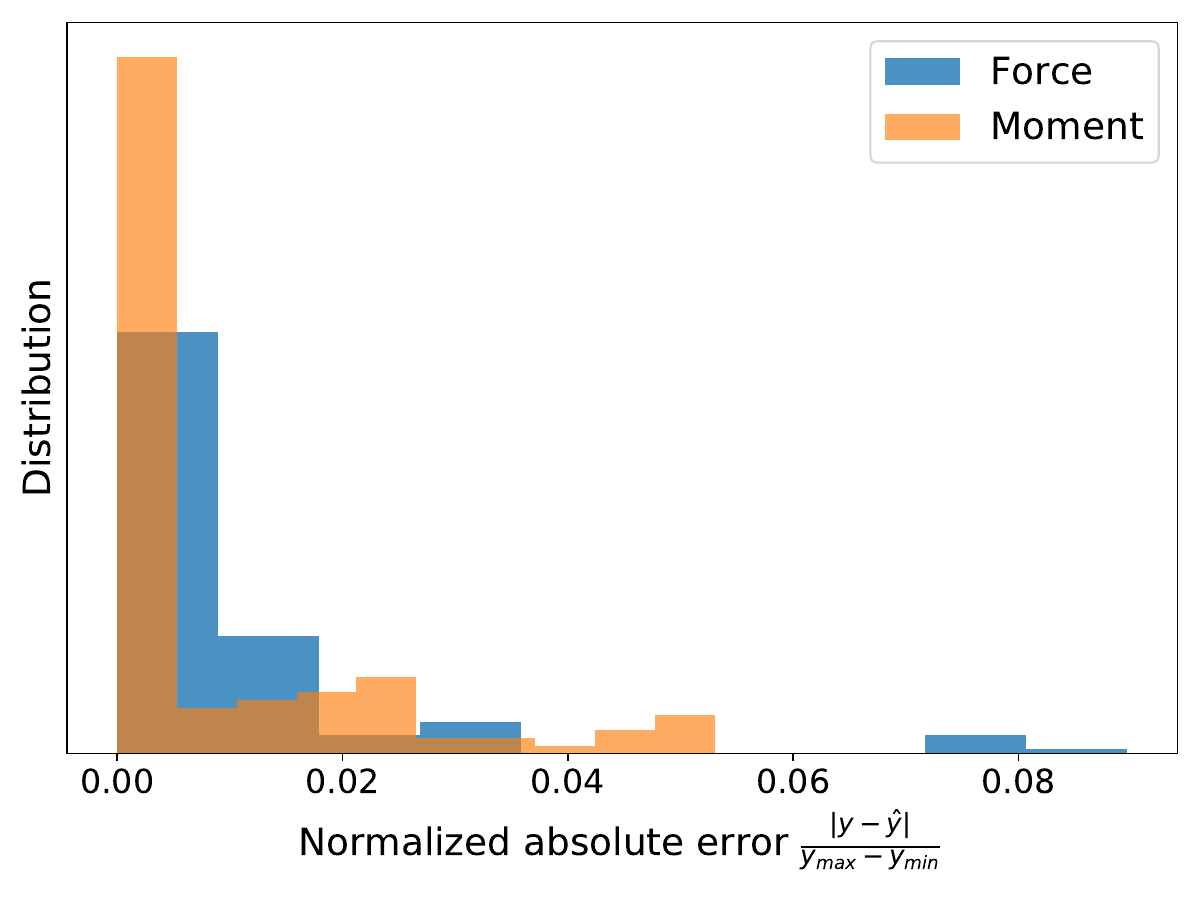}
\caption{}\label{fig:tension-torsion_loss_b}
\end{subfigure}
\caption{Rubber tension-torsion. (a) Training loss and number of effective parameters of the three neural networks . (b) Histogram of the normalized errors between the predicted and experimentally measured forces and moments.}
\label{fig:tension-torsion_loss}
\end{figure}

\begin{figure}
\centering
\includegraphics[scale=0.4]{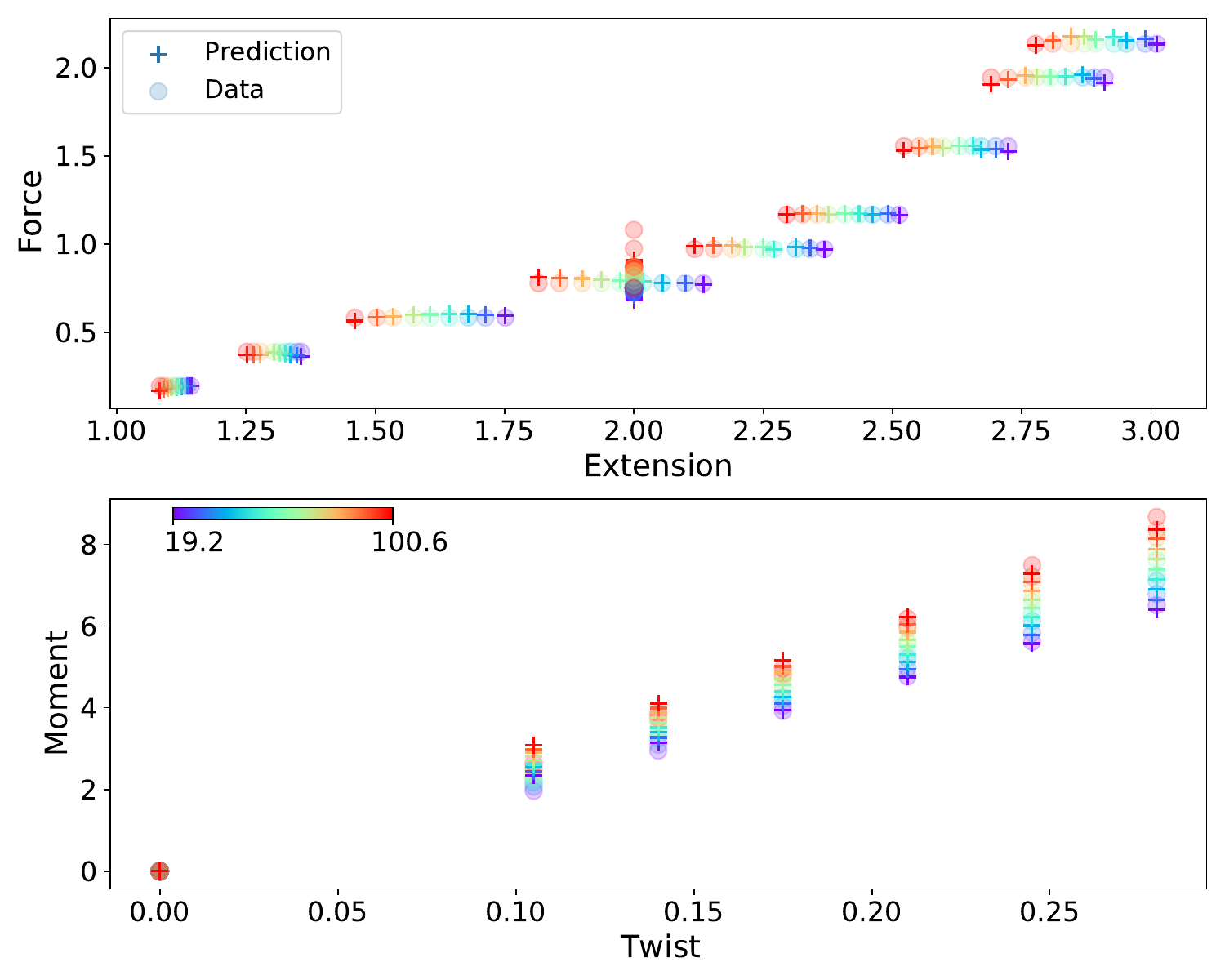}
\caption{Rubber tension-torsion force and moment predictions (crosses) and experimental data (circles) for temperatures from 19.2 to 100.6 $^\circ$C.}
\label{fig:tension-torsion}
\end{figure}

\fref{fig:tension-torsion_loss_a} shows the progress of the sparsification procedure and the loss function during training.
The descent of the loss is smooth which aids in the sparsification.
The purely mechanical term is eliminated (0 parameters) as redundant while
4 parameters for coupled mechanical $\psi_1$ and 15 for the thermal component $\phi_1$ remain.
The errors between the predicted response and the experimental data are generally less than 4\%, as shown in \fref{fig:tension-torsion_loss_b}.
Furthermore, \fref{fig:tension-torsion} shows the pointwise data and the resultants obtained with the fitted potential.

\subsubsection{Uniaxial tension of pig tissue} \label{sec:zhang}

Zhang \etal  \cite{zhang2018temperature} collected tension data of pig ear tissue over a range of temperatures to guide cryosurgery practice.
Since only tensile data was provided, we employed a mean squared loss on the non-zero stress component of the Cauchy stress
\begin{equation}
\Lc = \frac{1}{N_D} \sum_{i=1}^{N_D} \big\| [ \sigmab_i - \hat{\sigmab}(\Bb_i, \abstemperature_i) ]_{11} \big\|^{2}
\end{equation}
with an additional term for the $L^{0}$ regularization.
We assume that the deformation is incompressible, i.e.\ $J_e = 1$.
To account for the incompressibility constraint we assume the adiabatic thermoelastic split $J = \det \Fb = J_\abstemperature J_e = J_\abstemperature$ from  \eref{eq:pisterSplit}.
Then, we assert $\overline{\Psi}(I_{1},I_{2},J,\abstemperature)$ to be the approximation of an unconstrained free energy function with a neural network.
We enforce the incompressibility constraint $J_e = 1$ by introducing the Lagrange multiplier $p$ in
\begin{equation}
\begin{aligned}
\Psi(I_{1},I_{2},J,\abstemperature) = \overline{\Psi}(I_{1},I_{2},J,\abstemperature) + (p+n) (J_e-1) \ ,
\end{aligned}
\end{equation}
where $n$ is used to set the stress response to be $\Sb=\mathbf{0}$ at $\Fb = \Ib$ and $\abstemperature=0$.

Using the following derivatives
\begin{equation}
\frac{\partial I_{1}}{\partial \Cb} = \Ib, \quad \frac{\partial I_{2}}{\partial \Cb} = I_{1} \Ib - \Cb,\quad
\frac{\partial I_{3}}{\partial \Cb} = J^{2} \Cb^{-1} = J_\abstemperature^{2} J_e^{2} \Cb^{-1}, \quad
\frac{\partial J}{\partial I_{3}} = \frac{1}{2 J} = \frac{1}{2 J_\abstemperature J_e}, \quad  \frac{\partial J_e}{\partial J} = \frac{1}{J_\abstemperature} \ ,
\end{equation}
and \eref{eq:cauchy_stress},  we arrive at:
\begin{equation}
\begin{aligned}
\hat{\sigmab} &=   \frac{2}{J} \Fb \left( \frac{\partial \overline{\Psi}}{\partial I_{1}}\Ib + \frac{\partial \overline{\Psi}}{\partial I_{2}} \left[ I_{1} \Ib - \Cb \right] + \frac{\partial \overline{\Psi}}{\partial J}\frac{J_\abstemperature}{2} \Cb^{-1}\right) \Fb^{T} +  (p+n) \Ib \ .
\end{aligned}
\end{equation}
The stress normalization constant $n$ is defined as
\begin{equation}
\begin{aligned}
n = \left. \left(-2  \frac{\partial \overline{\Psi}}{\partial I_{1}}   -  \frac{\partial \overline{\Psi}}{\partial J}\right)\right\rvert_{\Fb_{e}=\Ib, \abstemperature=0}.
\end{aligned}
\end{equation}
For uniaxial tension under incompressibility, the deformation gradient and the Cauchy stress reduce to
\begin{equation}
\Fb = \text{diag}\Bigl\{ \lambda, \sqrt{\frac{J_\abstemperature}{\lambda}}, \sqrt{\frac{J_\abstemperature}{\lambda}}\Bigr\} \quad \text{and} \quad \sigmab = \text{diag}\lbrace \sigma_{11}, 0,0 \rbrace.
\end{equation}
The conditions $\sigma_{22}=\sigma_{33}=0$ restrict the Lagrange multiplier to be
\begin{equation}
p = - \frac{2}{\sqrt{I_{3}} \lambda} \left[ \frac{\partial \overline{\Psi}}{\partial I_{1}} + \frac{\partial \overline{\Psi}}{\partial I_{2}} \left(  \lambda^{2} + \frac{2 \sqrt{I_{3}}}{\lambda} - \frac{I_{3}}{\lambda}\right) + \frac{\partial \overline{\Psi}}{\partial J} \frac{\lambda}{2 \sqrt{I_{3}}}\right] -n \ .
\end{equation}
The non-zero, uniaxial component of the Cauchy stress is then given by
\begin{equation}
\begin{aligned}
\hat{\sigma}_{11}
&=  \frac{2}{\sqrt{I_{3}}} \lambda^{2} \left( \frac{\partial \overline{\Psi}}{\partial I_{1}} + \frac{\partial \overline{\Psi}}{\partial I_{2}}   \frac{2 \sqrt{I_{3}}}{\lambda}  + \frac{\partial \overline{\Psi}}{\partial J} \frac{\sqrt{I_{3}}}{2}  \frac{1}{\lambda^{2}}\right) + p + n \ .
\end{aligned}
\end{equation}

Due to the limited data, we validate the output of our models by plotting the response for equibiaxial tension under different temperatures.
Incompressible equibiaxial tension is defined by
\begin{equation}
\Fb = \text{diag}\Bigl\{ \lambda, \lambda, \frac{J_\abstemperature}{\lambda^{2}}\Bigr\} \quad \text{and} \quad \sigmab= \text{diag}\lbrace T_{11}, T_{22},0 \rbrace \ ,
\end{equation}
with the invariants
\begin{equation}
\begin{aligned}
I_{1} = 2 \lambda^{2} + \frac{J^{2}_\abstemperature}{\lambda^{4}} = 2 \lambda^{2} + \frac{I_{3}}{\lambda^{4}}, \quad I_{2}= \lambda^{2} + 2 \frac{I_{3}}{\lambda^{2}} \ .
\end{aligned}
\end{equation}
Enforcing $\sigma_{33}=0$ sets the Lagrange multiplier to
\begin{equation}
\begin{aligned}
p = - \frac{2 \sqrt{I_{3}}}{\lambda^{4}} \left( \frac{\partial \overline{\Psi}}{\partial I_{1}} + \frac{\partial \overline{\Psi}}{\partial I_{2}}  2 \lambda^{2} + \frac{\partial \overline{\Psi}}{\partial J} \frac{\lambda^{2}}{2 \sqrt{I_{3}}} \right) - n \ .
\end{aligned}
\end{equation}
Therefore the predicted equibiaxial stresses are
\begin{equation}
\begin{aligned}
\hat{\sigma}_{11} = \hat{\sigma}_{22} = \frac{2 \lambda^{2}}{\sqrt{I_{3}}}  \left( \frac{\partial \overline{\Psi}}{\partial I_{1}} + \frac{\partial \overline{\Psi}}{\partial I_{2}} \left[  \lambda^{2} + \frac{I_{3}}{\lambda^{4}} \right] +  \frac{\partial \overline{\Psi}}{\partial J} \frac{\sqrt{I_{3}}}{2} \frac{1}{\lambda^{2}}\right) + (p+n) \ .
\end{aligned}
\end{equation}

\begin{figure}
\centering
\begin{subfigure}[b]{0.5\textwidth}
\centering
\includegraphics[scale=0.35]{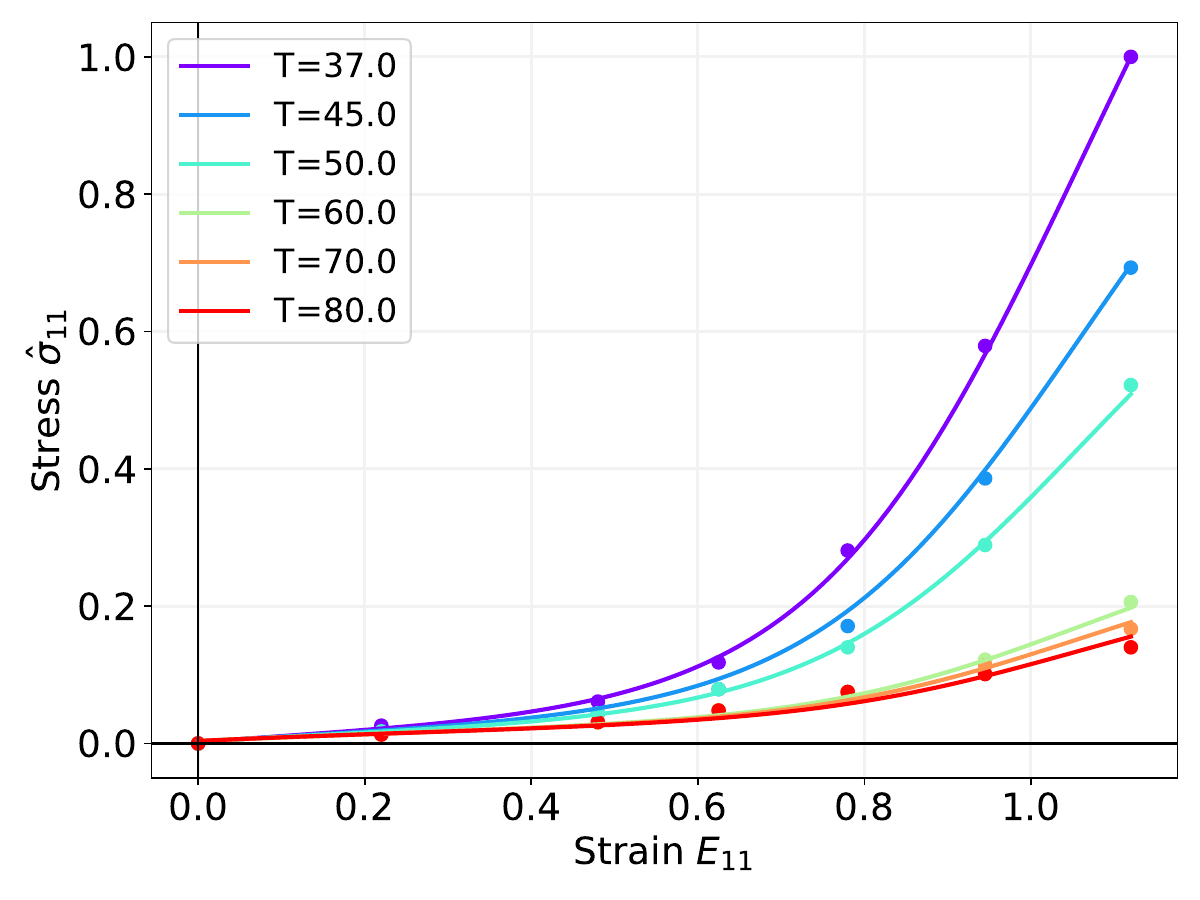}
\caption{}\label{fig:tension_a}
\end{subfigure}%
\begin{subfigure}{0.5\textwidth}
\centering
\includegraphics[scale=0.35]{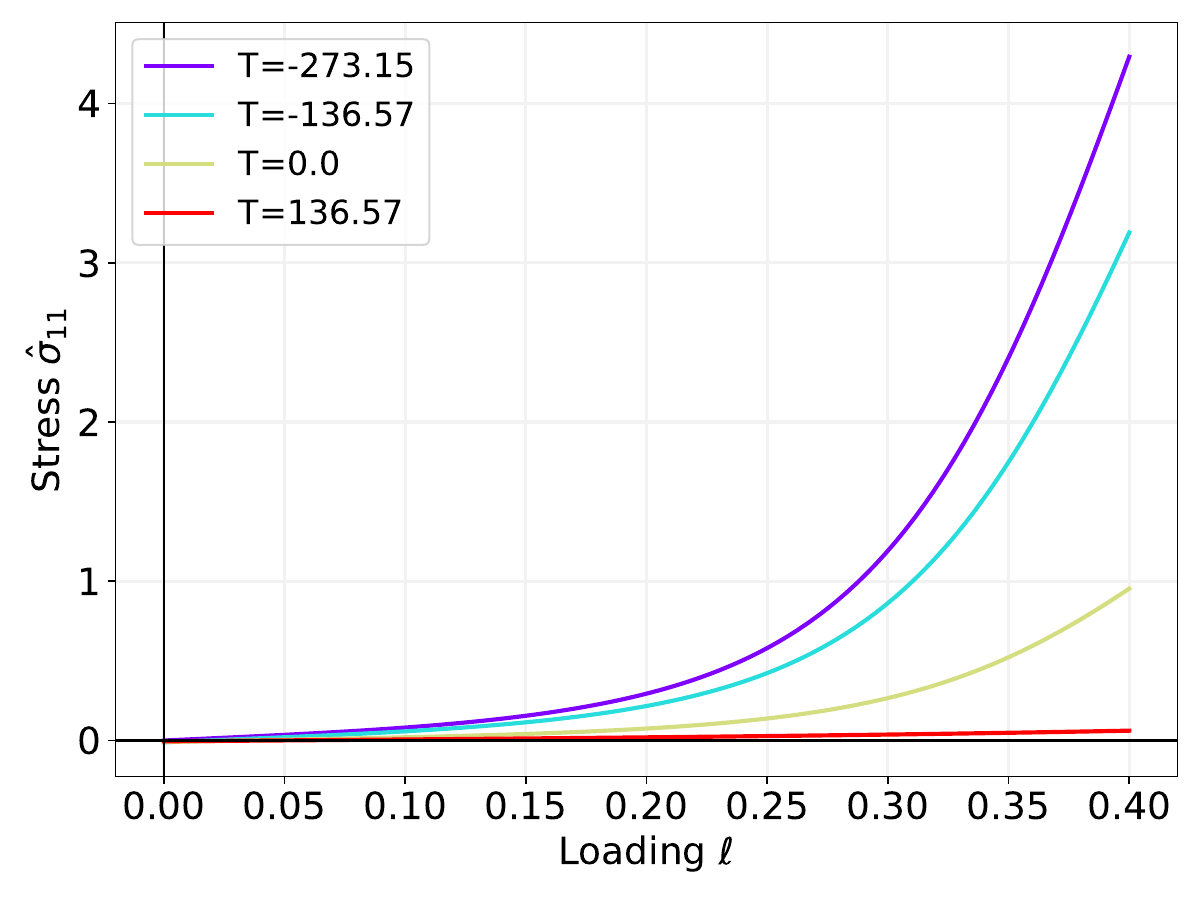}
\caption{}\label{fig:tension_b}
\end{subfigure}
\begin{subfigure}[b]{0.5\textwidth}
\centering
\includegraphics[scale=0.35]{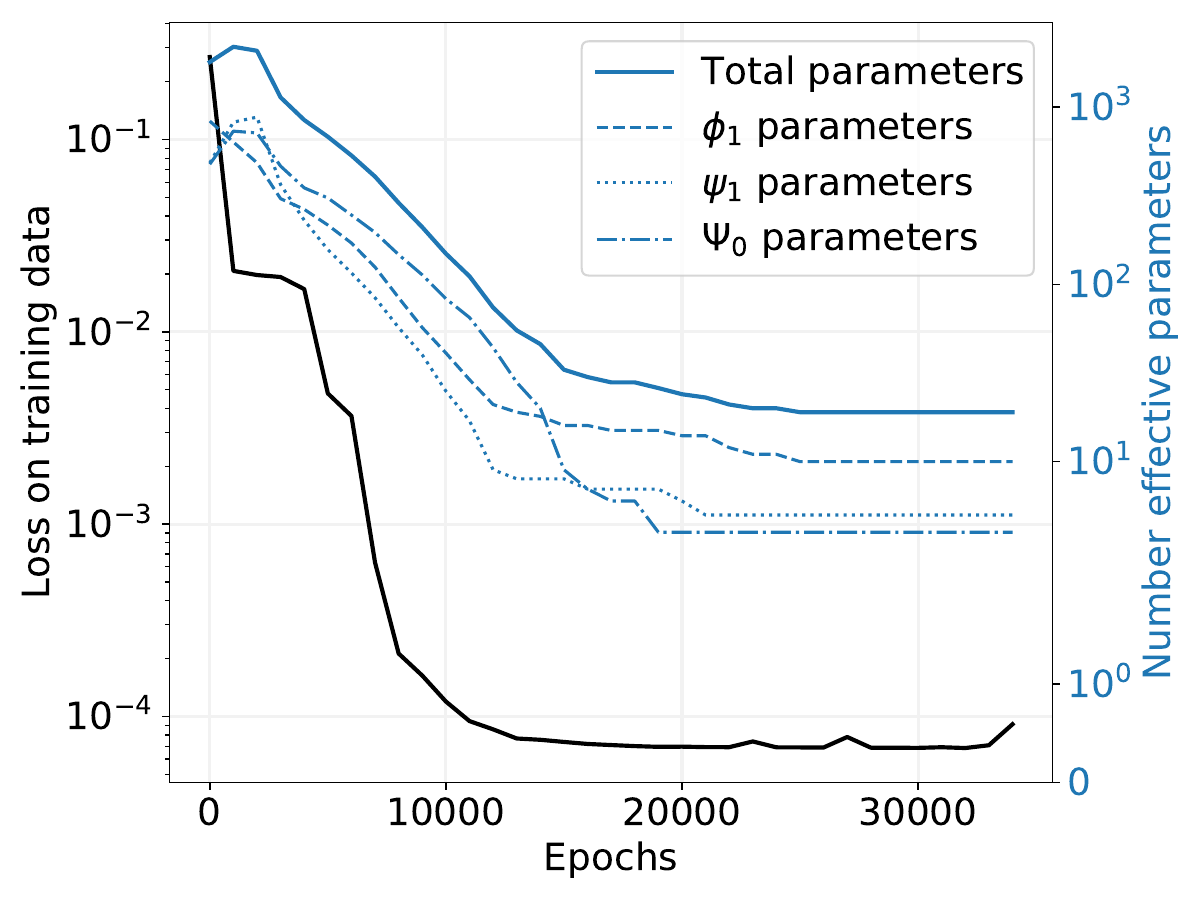}
\caption{}\label{fig:tension_c}
\end{subfigure}%
\begin{subfigure}{0.5\textwidth}
\centering
\includegraphics[scale=0.35]{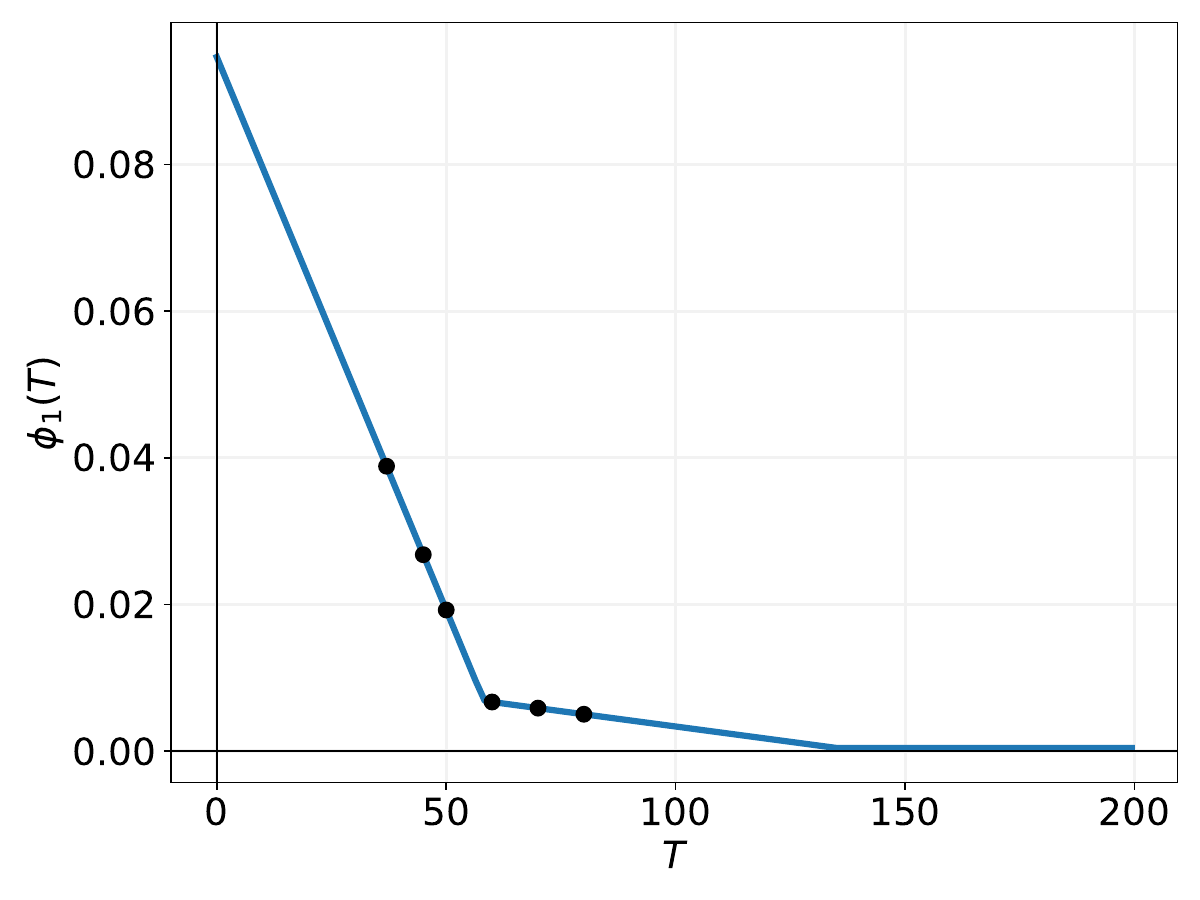}
\caption{}\label{fig:tension_d}
\end{subfigure}

\caption{Uniaxial tension of pig tissue. Temperatures are given in degrees Celsius ($^\circ$C). (a) Uniaxial tension data and predicted responses. (b) Biaxial response of calibrated model. (c) Loss on training data and number of effective parameters of the three networks. (d) Response of temperature-dependent potential $\phi_{i}$ over a large range of temperatures far exceeding the training data in temperature space (black dots). }
\label{fig:tension}
\end{figure}

The results shown in \fref{fig:tension_a} demonstrate that the model represents the data well while also predicting plausible responses for equibiaxial tension at various temperatures, see \fref{fig:tension_b}.
Using $L^{0}$ regularization, the total sum of trainable parameters of the models reaches around 20 after training, as seen in \fref{fig:tension_c}.
Furthermore, the functional form of the temperature-dependent strain energy component $\phi_{1}(T)$, as plotted in \fref{fig:tension_d} highlights that the model seems to be able to generalize plausibly far outside the range of temperatures in the training data.

\subsubsection{Uniaxial tension of carbon-filled black rubber} \label{sec:fu}
Lastly, we use the proposed framework to find a constitutive model for uniaxial tension data of carbon-filled black rubber presented in Fu \etal \cite{fu2021ability}. Hence, similarly to the previous case, we use $L^{0}$ regularized neural networks to fit the observed stress component $\sigma_{11}$ from a dataset of stretches $\lambda$ and uniaxial stress components at various temperatures.

\begin{figure}
\centering
\includegraphics[scale=0.4]{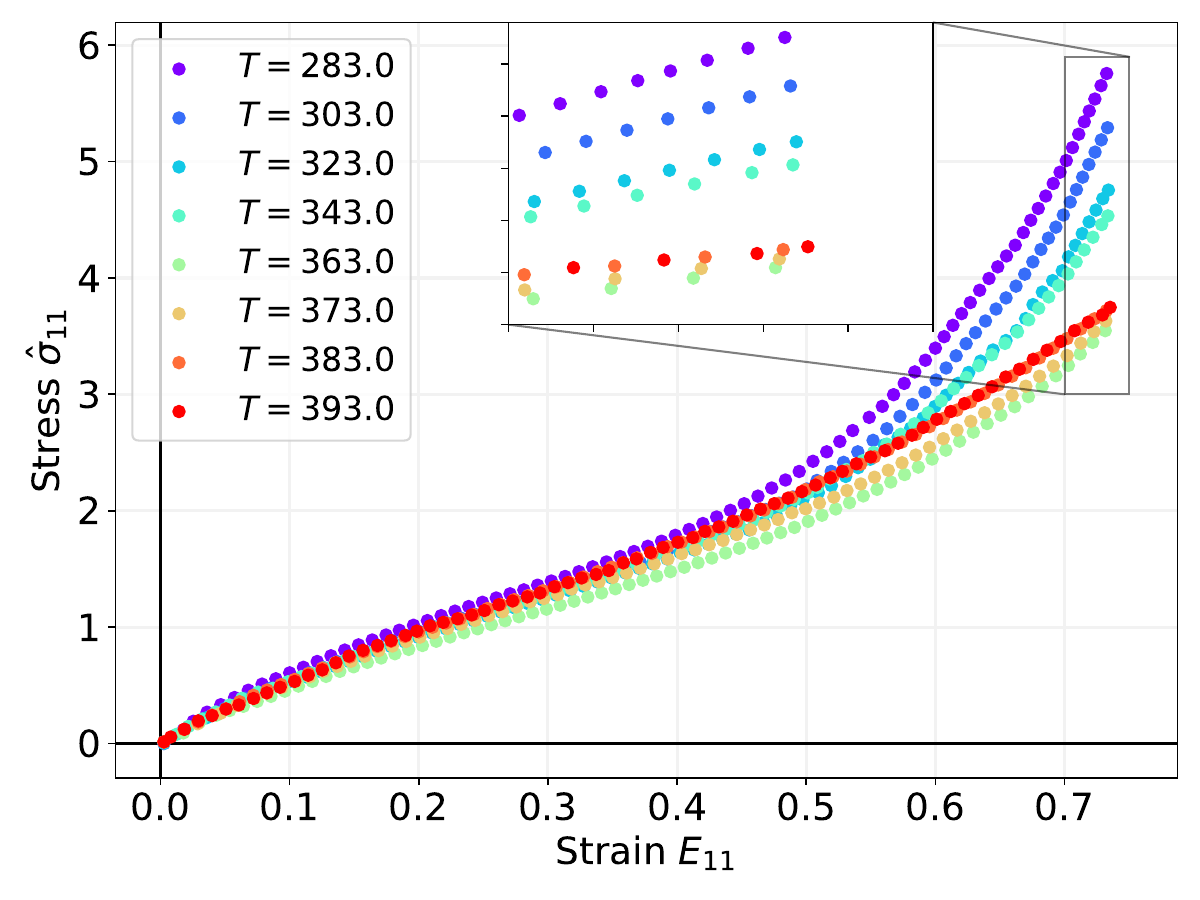}
\caption{Carbon-filled rubber dataset with inversion of stress-strain behavior with increasing temperatures. Temperatures are in Kelvin (K).}
\label{fig:raw_data_inversion}
\end{figure}

We explore this additional dataset because of an interesting temperature-dependent phenomenon.
Looking at the raw data in \fref{fig:raw_data_inversion}, we can see that in the range of $283~\text{K}<T<363~\text{K}$ the stress seems to decrease with increasing temperature, similar to the observations made for the pig tissue data in the previous section.
However, with a further increase in temperature, there appears to be an inversion in the behavior, i.e. stress increases with increasing temperature.
This phenomenon might be linked to a form of \emph{thermoelastic inversion}, c.f. Refs.
\cite{flory1953principles} and \cite{anthony1942equations}, which complicates the constitutive modeling process.

\begin{figure}
\centering
\begin{subfigure}[b]{0.5\textwidth}
\centering
\includegraphics[scale=0.35]{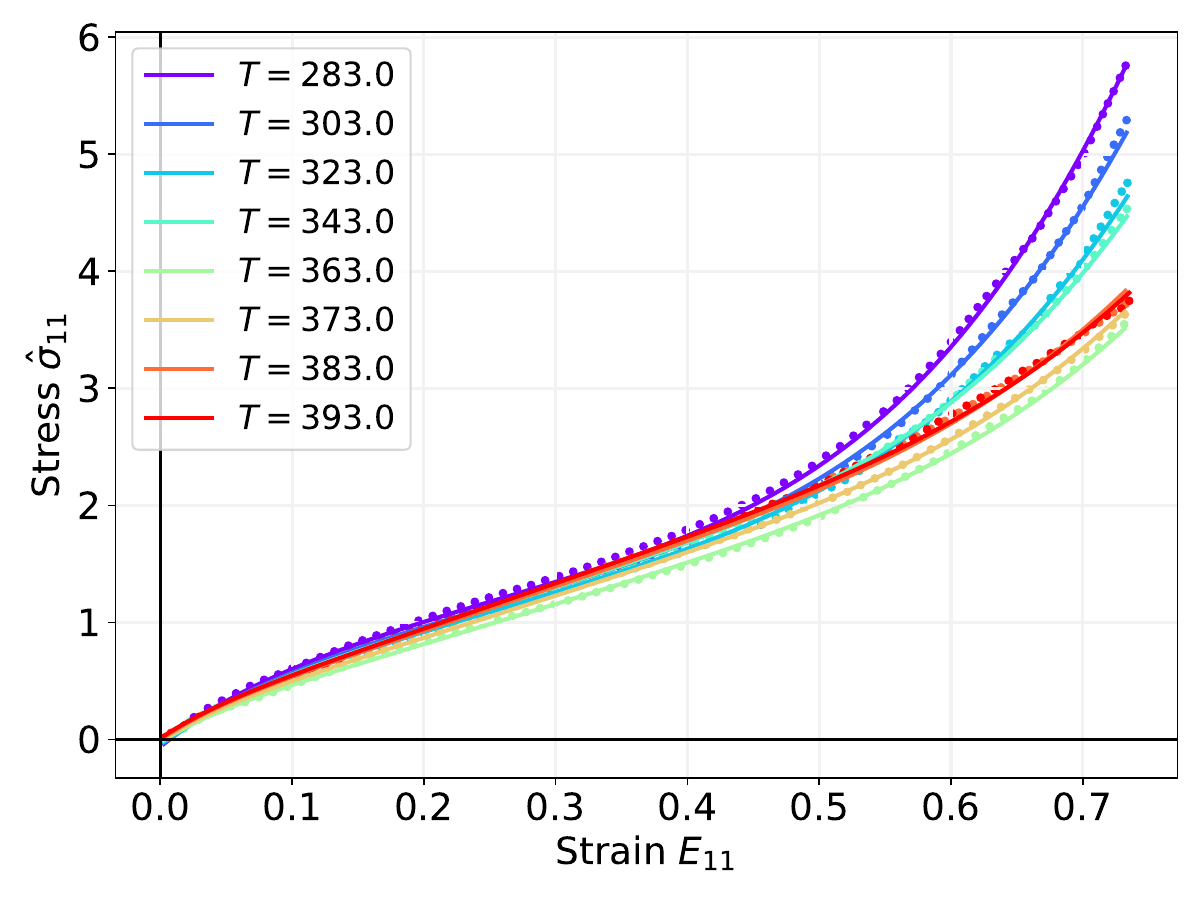}
\caption{}
\label{fig:tension_inversion_a}
\end{subfigure}%
\begin{subfigure}{0.5\textwidth}
\centering
\includegraphics[scale=0.35]{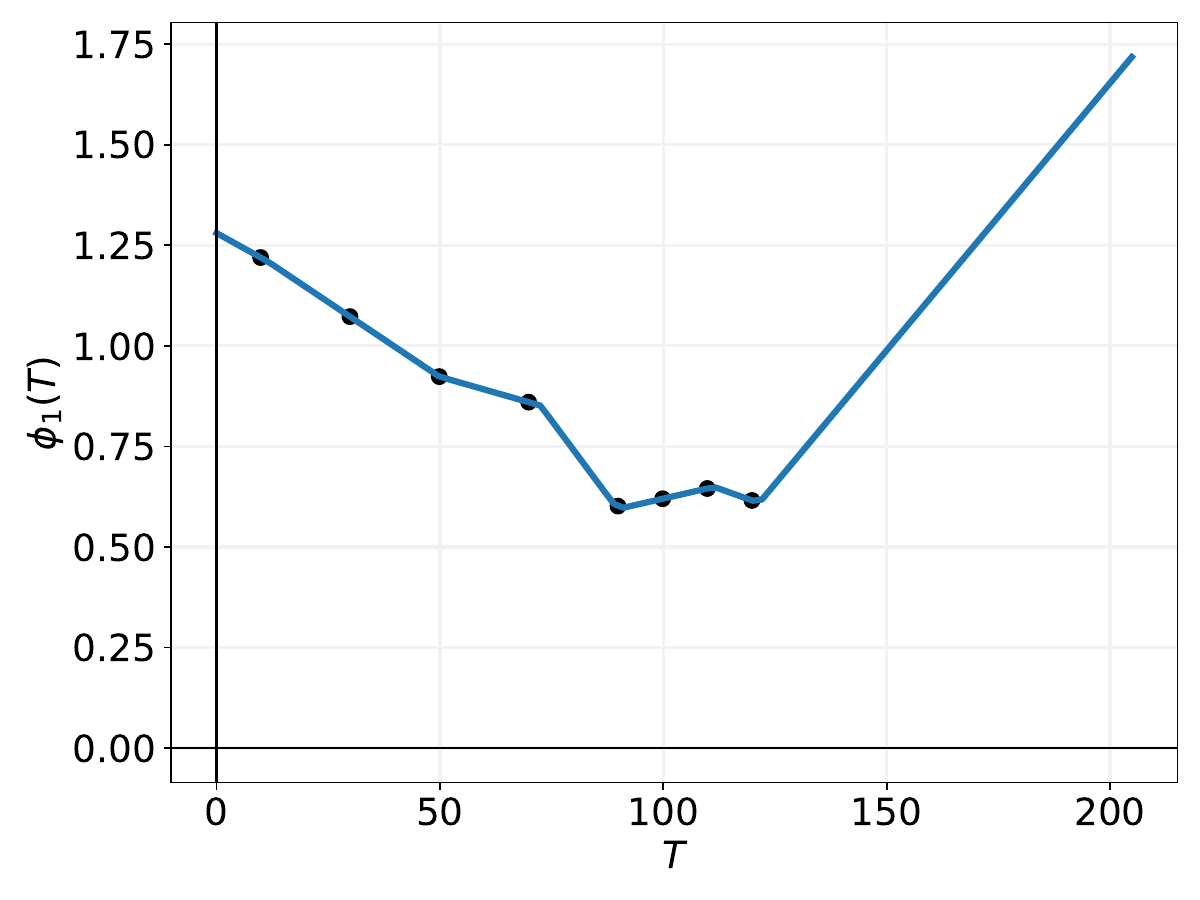}
\caption{}
\label{fig:tension_inversion_b}
\end{subfigure}
\begin{subfigure}{0.5\textwidth}
\centering
\includegraphics[scale=0.35]{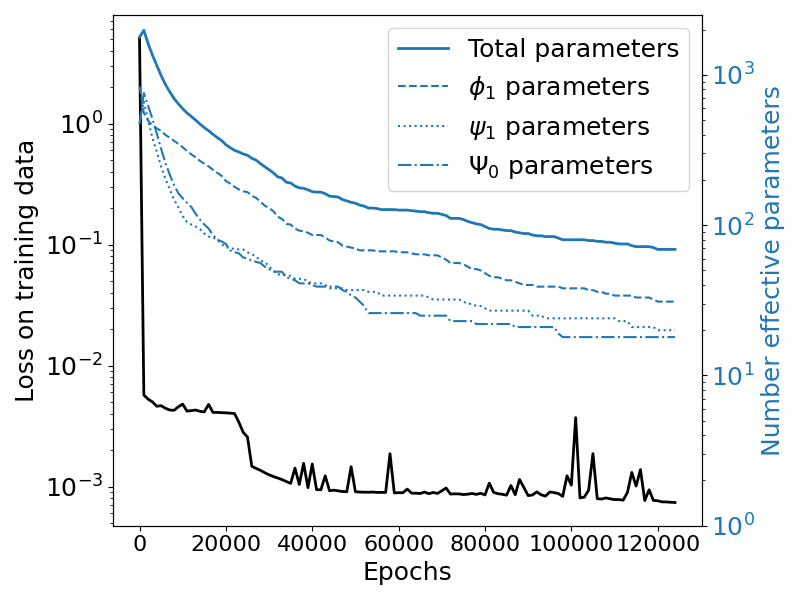}
\caption{}
\label{fig:tension_inversion_c}
\end{subfigure}%
\begin{subfigure}{0.5\textwidth}
\centering
\includegraphics[scale=0.35]{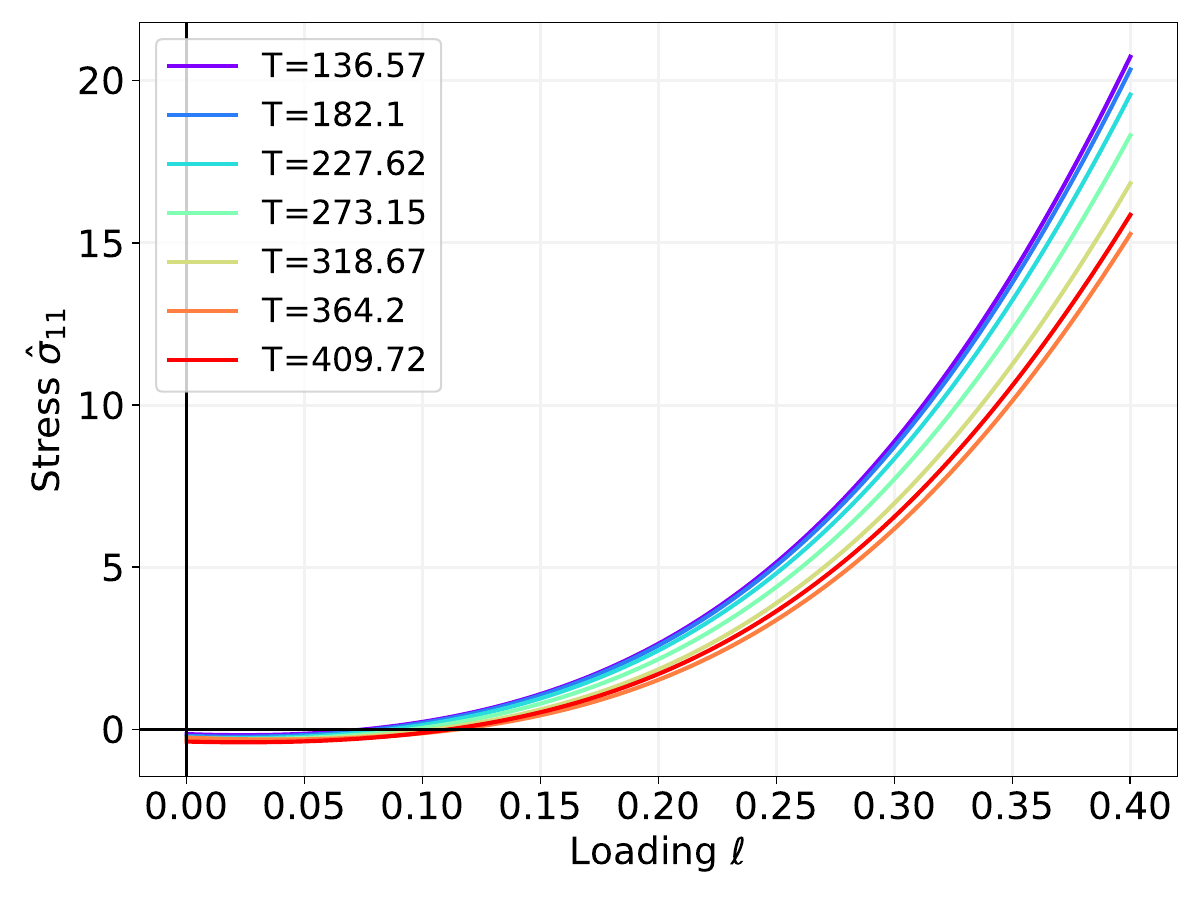}
\caption{}
\label{fig:tension_inversion_d}
\end{subfigure}
\caption{Results for the carbon-filled rubber dataset. Temperatures are given in Kelvin. (a) Uniaxial tension data and predicted responses. (b) Response of temperature-dependent potential $\phi_{i}$ over range of temperatures and training data in temperature space (black dots). (c) Loss on training data and number of effective parameters of the three networks. (d) Biaxial response of calibrated model.
}
\label{fig:tension_inversion}
\end{figure}

The predicted response of the calibrated model of the presented framework is shown in \fref{fig:tension_inversion_a}.
We can see that the response is in good agreement with the experimental data and captures the stress inversion.
This is highlighted by the response of the temperature-dependent potential $\phi_{i}$ shown in \fref{fig:tension_inversion_b}.
Here the black dots indicate available temperature measurements and the blue line is the $\phi_{i}$ response function.
The training loss and the number of effective parameters due to the regularization of the networks are shown in \fref{fig:tension_inversion_c}.
The relatively low number of final parameters appears to improve generalization since the response of the model in equibiaxial loading is plausible and also displays the thermal inversion, see
\fref{fig:tension_inversion_d}.

\section{Conclusion} \label{sec:conclusion}

We presented a complete representation of physics-constrained machine learning of thermoelastic stress behavior based on a thermomechanically constrained free energy function and demonstrated the ability of its neural network implementation to accurately model both synthetic and experimental data.
The model is capable of representing non-self-similar temperature dependence, general thermal expansion, thermal softening, and thermal inversion.
The series representation is amenable to both term-wise and parameter-wise sparsification that imparts a degree of generalization and interpretability to the resulting models.

The formalism is directly extensible to anisotropic thermo-hyperelasticity through the inclusion of additional invariants associated with a structure tensor that characterizes the symmetry group and use of the isotropization theorem.
As we have shown in \cref{fuhg2022learning}, these symmetries can be learned, as well as presupposed.

\section*{Acknowledgements}
PyTorch \cite{pytorch}  was used to implement the models and methods of this work.
This open-source software is gratefully acknowledged.
O.W. acknowledges the financial support provided by the Deutsche Forschungsgemeinschaft (DFG, German Research Foundation, project number 492770117).
R.E.J. and D.T.S. were supported by the Laboratory Directed Research and Development program at Sandia National Laboratories, a multimission laboratory managed and operated by National Technology and Engineering Solutions of Sandia, LLC, a wholly owned subsidiary of Honeywell International, Inc., for the U.S. Department of Energy's National Nuclear Security Administration under contract DE-NA-0003525.
This paper describes objective technical results and analysis. Any subjective views or opinions that might be expressed in the paper do not necessarily represent the views of the U.S. Department of Energy or the United States Government.

\bibliographystyle{unsrt}

\begin{thebibliography}{10}

\bibitem{klein2022polyconvex}
Dominik~K Klein, Mauricio Fern{\'a}ndez, Robert~J Martin, Patrizio Neff, and
Oliver Weeger.
\newblock Polyconvex anisotropic hyperelasticity with neural networks.
\newblock {\em Journal of the Mechanics and Physics of Solids}, 159:104703,
2022.

\bibitem{chen2022polyconvex}
Peiyi Chen and Johann Guilleminot.
\newblock Polyconvex neural networks for hyperelastic constitutive models: A
rectification approach.
\newblock {\em Mechanics Research Communications}, 125:103993, 2022.

\bibitem{tac2022data}
Vahidullah Tac, Francisco~Sahli Costabal, and Adrian~B Tepole.
\newblock Data-driven tissue mechanics with polyconvex neural ordinary
differential equations.
\newblock {\em Computer Methods in Applied Mechanics and Engineering},
398:115248, 2022.

\bibitem{fuhg2024stress}
Jan Fuhg, Nikolaos Bouklas, and Reese Jones.
\newblock Stress representations for tensor basis neural networks: alternative
formulations to finger-rivlin-ericksen.
\newblock {\em Journal of Computing and Information Science in Engineering},
pages 1--39, 2024.

\bibitem{zlatic2023incompressible}
Martin Zlati{\'c} and Marko {\v{C}}anadija.
\newblock Incompressible rubber thermoelasticity: a neural network approach.
\newblock {\em Computational Mechanics}, 71(5):895--916, 2023.

\bibitem{swiler2020survey}
Laura~P Swiler, Mamikon Gulian, Ari~L Frankel, Cosmin Safta, and John~D
Jakeman.
\newblock A survey of constrained gaussian process regression: Approaches and
implementation challenges.
\newblock {\em Journal of Machine Learning for Modeling and Computing}, 1(2),
2020.

\bibitem{YVONNET20092723}
J.~Yvonnet, D.~Gonzalez, and Q.-C. He.
\newblock Numerically explicit potentials for the homogenization of nonlinear
elastic heterogeneous materials.
\newblock {\em Computer Methods in Applied Mechanics and Engineering},
198(33):2723--2737, 2009.

\bibitem{coleman1963thermodynamics}
Bernard~D Coleman and Walter Noll.
\newblock The thermodynamics of elastic materials with heat conduction and
viscosity.
\newblock {\em Archive for Rational Mechanics and Analysis}, 13:167--178, 1963.

\bibitem{truesdell2004non}
Clifford Truesdell and Walter Noll.
\newblock {\em The non-linear field theories of mechanics}.
\newblock Springer, 2004.

\bibitem{rivlin1986reflections}
Ronald~S Rivlin.
\newblock Reflections on certain aspects of thermomechanics.
\newblock In {\em Collected Papers of RS Rivlin: Volume I and II}, pages
430--463. Springer, 1986.

\bibitem{lu1975decomposition}
SCH Lu and KS~Pister.
\newblock Decomposition of deformation and representation of the free energy
function for isotropic thermoelastic solids.
\newblock {\em International Journal of Solids and Structures},
11(7-8):927--934, 1975.

\bibitem{lee1969elastic}
EH~Lee.
\newblock Elastic-plastic deformation at finite strains.
\newblock {\em Journal of Applied Mechanics}, 36(1):1--6, 1969.

\bibitem{vujovsevic2002finite}
L~Vujo{\v{s}}evi{\'c} and VA~Lubarda.
\newblock Finite-strain thermoelasticity based on multiplicative decomposition
of deformation gradient.
\newblock {\em Theoretical and applied mechanics}, (28-29):379--399, 2002.

\bibitem{lubarda2004constitutive}
Vlado~A Lubarda.
\newblock Constitutive theories based on the multiplicative decomposition of
deformation gradient: Thermoelasticity, elastoplasticity, and biomechanics.
\newblock {\em Appl. Mech. Rev.}, 57(2):95--108, 2004.

\bibitem{bouteiller2023complete}
Paul Bouteiller.
\newblock Complete finite-strain isotropic thermo-elasticity.
\newblock {\em European Journal of Mechanics-A/Solids}, 100:105017, 2023.

\bibitem{franke2018energy}
M~Franke, A~Janz, Mark Schiebl, and Peter Betsch.
\newblock An energy momentum consistent integration scheme using a
polyconvexity-based framework for nonlinear thermo-elastodynamics.
\newblock {\em International Journal for Numerical Methods in Engineering},
115(5):549--577, 2018.

\bibitem{bonet2021first}
Javier Bonet, Chun~Hean Lee, Antonio~J Gil, and Ataollah Ghavamian.
\newblock A first order hyperbolic framework for large strain computational
solid dynamics. part iii: Thermo-elasticity.
\newblock {\em Computer Methods in Applied Mechanics and Engineering},
373:113505, 2021.

\bibitem{armero1992new}
F~Armero and JC~Simo.
\newblock A new unconditionally stable fractional step method for non-linear
coupled thermomechanical problems.
\newblock {\em International Journal for numerical methods in Engineering},
35(4):737--766, 1992.

\bibitem{holzapfel1996entropy}
Gerhard~A Holzapfel and JC~Simo.
\newblock Entropy elasticity of isotropic rubber-like solids at finite strains.
\newblock {\em Computer Methods in applied mechanics and engineering},
132(1-2):17--44, 1996.

\bibitem{tamma1997computational}
Kumar~K Tamma and Raju~R Namburu.
\newblock Computational approaches with applications to non-classical and
classical thermomechanical problems.
\newblock 1997.

\bibitem{casey1998characterization}
J~Casey and S~Krishnaswamy.
\newblock A characterization of internally constrained thermoelastic materials.
\newblock {\em Mathematics and Mechanics of Solids}, 3(1):71--89, 1998.

\bibitem{casey1998elastic}
James Casey.
\newblock On elastic-thermo-plastic materials at finite deformations.
\newblock {\em International Journal of Plasticity}, 14(1-3):173--191, 1998.

\bibitem{casey2011nonlinear}
James Casey.
\newblock Nonlinear thermoelastic materials with viscosity, and subject to
internal constraints: a classical continuum thermodynamics approach.
\newblock {\em Journal of Elasticity}, 104:91--104, 2011.

\bibitem{miehe1995entropic}
C~Miehe.
\newblock Entropic thermoelasticity at finite strains. aspects of the
formulation and numerical implementation.
\newblock {\em Computer Methods in Applied Mechanics and Engineering},
120(3-4):243--269, 1995.

\bibitem{horgan2003finite}
Cornelius~O Horgan and Giuseppe Saccomandi.
\newblock Finite thermoelasticity with limiting chain extensibility.
\newblock {\em Journal of the Mechanics and Physics of Solids},
51(6):1127--1146, 2003.

\bibitem{doi1988theory}
Masao Doi and Samuel~Frederick Edwards.
\newblock {\em The theory of polymer dynamics}, volume~73.
\newblock oxford university press, 1988.

\bibitem{treloar1975physics}
LR~G Treloar.
\newblock The physics of rubber elasticity.
\newblock 1975.

\bibitem{heinrich2003thermoelasticity}
G~Heinrich, M~Kaliske, M~Kl{\"u}ppel, JE~Mark, E~Straube, and Thomas~A Vilgis.
\newblock The thermoelasticity of rubberlike materials and related constitutive
laws.
\newblock {\em Journal of Macromolecular Science, Part A}, 40(1):87--93, 2003.

\bibitem{kimmer2007continuum}
CJ~Kimmer and RE~Jones.
\newblock Continuum constitutive models from analytical free energies.
\newblock {\em Journal of Physics: Condensed Matter}, 19(32):326207, 2007.

\bibitem{linka2023new}
Kevin Linka and Ellen Kuhl.
\newblock A new family of constitutive artificial neural networks towards
automated model discovery.
\newblock {\em Computer Methods in Applied Mechanics and Engineering},
403:115731, 2023.

\bibitem{amos2017input}
Brandon Amos, Lei Xu, and J~Zico Kolter.
\newblock Input convex neural networks.
\newblock In {\em International Conference on Machine Learning}, pages
146--155. PMLR, 2017.

\bibitem{richter2021input}
Jack Richter-Powell, Jonathan Lorraine, and Brandon Amos.
\newblock Input convex gradient networks.
\newblock {\em arXiv preprint arXiv:2111.12187}, 2021.

\bibitem{makkuva2020optimal}
Ashok Makkuva, Amirhossein Taghvaei, Sewoong Oh, and Jason Lee.
\newblock Optimal transport mapping via input convex neural networks.
\newblock In {\em International Conference on Machine Learning}, pages
6672--6681. PMLR, 2020.

\bibitem{chen2018optimal}
Yize Chen, Yuanyuan Shi, and Baosen Zhang.
\newblock Optimal control via neural networks: A convex approach.
\newblock {\em arXiv preprint arXiv:1805.11835}, 2018.

\bibitem{as2022mechanics}
Faisal As'ad, Philip Avery, and Charbel Farhat.
\newblock A mechanics-informed artificial neural network approach in
data-driven constitutive modeling.
\newblock {\em International Journal for Numerical Methods in Engineering},
123(12):2738--2759, 2022.

\bibitem{xu2021learning}
Kailai Xu, Daniel~Z Huang, and Eric Darve.
\newblock Learning constitutive relations using symmetric positive definite
neural networks.
\newblock {\em Journal of Computational Physics}, 428:110072, 2021.

\bibitem{klein2023parametrized}
Dominik~K Klein, Fabian~J Roth, Iman Valizadeh, and Oliver Weeger.
\newblock Parametrized polyconvex hyperelasticity with physics-augmented neural
networks.
\newblock {\em Data-Centric Engineering}, 4:e25, 2023.

\bibitem{kalina2024neural}
Karl~A Kalina, Philipp Gebhart, J{\"o}rg Brummund, Lennart Linden, WaiChing
Sun, and Markus K{\"a}stner.
\newblock Neural network-based multiscale modeling of finite strain
magneto-elasticity with relaxed convexity criteria.
\newblock {\em Computer Methods in Applied Mechanics and Engineering},
421:116739, 2024.

\bibitem{fuhg2022learning}
Jan~N Fuhg, Nikolaos Bouklas, and Reese~E Jones.
\newblock Learning hyperelastic anisotropy from data via a tensor basis neural
network.
\newblock {\em Journal of the Mechanics and Physics of Solids}, 168:105022,
2022.

\bibitem{fuhg2022machine}
Jan~N Fuhg, Lloyd van Wees, Mark Obstalecki, Paul Shade, Nikolaos Bouklas, and
Matthew Kasemer.
\newblock Machine-learning convex and texture-dependent macroscopic yield from
crystal plasticity simulations.
\newblock {\em Materialia}, 23:101446, 2022.

\bibitem{linden2023neural}
Lennart Linden, Dominik~K Klein, Karl~A Kalina, J{\"o}rg Brummund, Oliver
Weeger, and Markus K{\"a}stner.
\newblock Neural networks meet hyperelasticity: A guide to enforcing physics.
\newblock {\em Journal of the Mechanics and Physics of Solids}, page 105363,
2023.

\bibitem{silhavy2013mechanics}
Miroslav Silhavy.
\newblock {\em The mechanics and thermodynamics of continuous media}.
\newblock Springer Science \& Business Media, 2013.

\bibitem{ashcroft1976solid}
Neil~W Ashcroft and N~David Mermin.
\newblock {\em Solid state physics}.
\newblock Cengage Learning, 1976.

\bibitem{kolmogorov1957elements}
Andre\u{\i}~Nikolaevich Kolmogorov and Serge\u{\i}~Vasilevich Fomin.
\newblock {\em Elements of the theory of functions and functional analysis},
volume~1.
\newblock Courier Corporation, 1957.

\bibitem{bigoni2016spectral}
Daniele Bigoni, Allan~P Engsig-Karup, and Youssef~M Marzouk.
\newblock Spectral tensor-train decomposition.
\newblock {\em SIAM Journal on Scientific Computing}, 38(4):A2405--A2439, 2016.

\bibitem{griebel2023analysis}
Michael Griebel and Helmut Harbrecht.
\newblock Analysis of tensor approximation schemes for continuous functions.
\newblock {\em Foundations of Computational Mathematics}, pages 1--22, 2023.

\bibitem{louizos2017learning}
Christos Louizos, Max Welling, and Diederik~P Kingma.
\newblock Learning sparse neural networks through $ l\_0 $ regularization.
\newblock {\em arXiv preprint arXiv:1712.01312}, 2017.

\bibitem{fuhg2023extreme}
Jan~N Fuhg, Reese~E Jones, and Nikolaos Bouklas.
\newblock Extreme sparsification of physics-augmented neural networks for
interpretable model discovery in mechanics.
\newblock {\em arXiv preprint arXiv:2310.03652}, 2023.

\bibitem{ratku2022derivatives}
Antal Ratku and Dirk Neumann.
\newblock Derivatives of feed-forward neural networks and their application in
real-time market risk management.
\newblock {\em OR Spectrum}, pages 1--19, 2022.

\bibitem{fuhg2023modular}
Jan~Niklas Fuhg, Craig~M Hamel, Kyle Johnson, Reese Jones, and Nikolaos
Bouklas.
\newblock Modular machine learning-based elastoplasticity: Generalization in
the context of limited data.
\newblock {\em Computer Methods in Applied Mechanics and Engineering},
407:115930, 2023.

\bibitem{zhang2018temperature}
Jinao Zhang, Jeremy Hills, Yongmin Zhong, Bijan Shirinzadeh, Julian Smith, and
Chengfan Gu.
\newblock Temperature-dependent thermomechanical modeling of soft tissue
deformation.
\newblock {\em Journal of Mechanics in Medicine and Biology}, 18(08):1840021,
2018.

\bibitem{mohsin1987thermoelastic}
Mahmood~A Mohsin and Leslie~RG Treloar.
\newblock Thermoelastic measurements of some elastomers under extension and
torsion.
\newblock {\em Polymer}, 28(11):1893--1898, 1987.

\bibitem{fu2021ability}
Xintao Fu, Zepeng Wang, and Lianxiang Ma.
\newblock Ability of constitutive models to characterize the temperature
dependence of rubber hyperelasticity and to predict the stress-strain
behavior of filled rubber under different defor mation states.
\newblock {\em Polymers}, 13(3):369, 2021.

\bibitem{kingma2014adam}
Diederik~P Kingma and Jimmy Ba.
\newblock Adam: A method for stochastic optimization.
\newblock {\em arXiv preprint arXiv:1412.6980}, 2014.

\bibitem{rivlin1997collected}
Ronald~S Rivlin and Grigory~I Barenblatt.
\newblock {\em Collected papers of RS Rivlin}, volume~1.
\newblock Springer Science \& Business Media, 1997.

\bibitem{mie1903kinetischen}
Gustav Mie.
\newblock Zur kinetischen theorie der einatomigen k{\"o}rper.
\newblock {\em Annalen der Physik}, 316(8):657--697, 1903.

\bibitem{gruneisen1912theorie}
Eduard Gr{\"u}neisen.
\newblock Theorie des festen zustandes einatomiger elemente.
\newblock {\em Annalen der Physik}, 344(12):257--306, 1912.

\bibitem{rivlin1948large}
Ronald~S Rivlin.
\newblock Large elastic deformations of isotropic materials iv. further
developments of the general theory.
\newblock {\em Philosophical transactions of the royal society of London.
Series A, Mathematical and physical sciences}, 241(835):379--397, 1948.

\bibitem{rivlin1951large}
Ronald~S Rivlin and DW~Saunders.
\newblock Large elastic deformations of isotropic materials vii. experiments on
the deformation of rubber.
\newblock {\em Philosophical Transactions of the Royal Society of London.
Series A, Mathematical and Physical Sciences}, 243(865):251--288, 1951.

\bibitem{hartmann2001numerical}
S~Hartmann.
\newblock Numerical studies on the identification of the material parameters of
rivlin's hyperelasticity using tension-torsion tests.
\newblock {\em Acta Mechanica}, 148(1-4):129--155, 2001.

\bibitem{kirkinis2002extension}
E~Kirkinis and RW19214631045 Ogden.
\newblock On extension and torsion of a compressible elastic circular cylinder.
\newblock {\em Mathematics and Mechanics of Solids}, 7(4):373--392, 2002.

\bibitem{flory1953principles}
Paul~J Flory.
\newblock {\em Principles of polymer chemistry}.
\newblock Cornell university press, 1953.

\bibitem{anthony1942equations}
Robert~Louis Anthony, Ralph~Henry Caston, and Eugene Guth.
\newblock Equations of state for natural and synthetic rubber-like materials.
i. unaccelerated natural soft rubber.
\newblock {\em The Journal of Physical Chemistry}, 46(8):826--840, 1942.

\bibitem{pytorch}
Adam Paszke, Sam Gross, Francisco Massa, Adam Lerer, James Bradbury, Gregory
Chanan, Trevor Killeen, Zeming Lin, Natalia Gimelshein, Luca Antiga, Alban
Desmaison, Andreas Kopf, Edward Yang, Zachary DeVito, Martin Raison, Alykhan
Tejani, Sasank Chilamkurthy, Benoit Steiner, Lu~Fang, Junjie Bai, and Soumith
Chintala.
\newblock Pytorch: An imperative style, high-performance deep learning library.
\newblock In {\em Advances in Neural Information Processing Systems 32}, pages
8024--8035. Curran Associates, Inc., 2019.

\end{thebibliography}

\end{document}